\documentclass[aps,english,prl,floatfix,amsmath,superscriptaddress,tightenlines,twocolumn]{revtex4-2}
\usepackage{mathtools, amssymb}
\usepackage{graphicx}
\usepackage{tikz}
\usepackage{tikz-3dplot}
\usetikzlibrary{spy}
\usetikzlibrary{arrows.meta}
\usetikzlibrary{calc}
\usetikzlibrary{decorations.pathreplacing,calligraphy}
\usepackage{intcalc}
\usepackage[caption=false]{subfig}
\usepackage[shortlabels]{enumitem}
\usepackage{soul}
\usepackage[utf8]{inputenc}
\usepackage{xcolor}
\usepackage{amsthm}
\usepackage{float}
\usepackage{comment}
\usepackage{braket}

\newcommand{\tr}{\textrm{Tr}}

\newcommand{\eqfig}[2]{\vcenter{\hbox{\includegraphics[height=#1]{#2}}}}

\begin{document}

\title{Channeling quantum criticality}

\author{Yijian Zou}
\affiliation{Stanford Institute for Theoretical Physics, Stanford University, Stanford, CA 94305, USA}

\author{Shengqi Sang}
\affiliation{Perimeter Institute for Theoretical Physics, Waterloo, Ontario N2L 2Y5, Canada}
\affiliation{University of Waterloo, Waterloo, Ontario, N2L 3G1, Canada}

\author{Timothy H. Hsieh}
\affiliation{Perimeter Institute for Theoretical Physics, Waterloo, Ontario N2L 2Y5, Canada}
\begin{abstract}
We analyze the effect of decoherence, modelled by local quantum channels, on quantum critical states and we find universal properties of the resulting mixed state's entanglement, both between system and environment and within the system.  Renyi entropies exhibit volume law scaling with a subleading constant governed by a ``$g$-function'' in conformal field theory (CFT), allowing us to define a notion of renormalization group (RG) flow (or ``phase transitions'') between quantum channels.  We also find that the entropy of a subsystem in the decohered state has a subleading logarithmic scaling with subsystem size, and we relate it to correlation functions of boundary condition changing operators in the CFT.  Finally, we find that the subsystem entanglement negativity, a measure of quantum correlations within mixed states, can exhibit log scaling or area law based on the RG flow. When the channel corresponds to a marginal perturbation, the coefficient of the log scaling can change continuously with decoherence strength. We illustrate all these possibilities for the critical ground state of the transverse-field Ising model, in which we identify four RG fixed points of dephasing channels and verify the RG flow numerically.  Our results are relevant to quantum critical states realized on noisy quantum simulators, in which our predicted entanglement scaling can be probed via shadow tomography methods.         

\end{abstract}

\maketitle
\emph{Introduction --} 
Quantum devices have advanced significantly to the point of challenging  classical computers in sampling and simulation tasks  \cite{google2019supremacy,doi:10.1126/science.abe8770, xanadu, zhang, bernien}. In the context of quantum many-body physics, they are promising experimental platforms for realizing long-range entangled quantum phases of matter including topological order \cite{doi:10.1126/science.abi8794,doi:10.1126/science.abi8378} and quantum critical states \cite{zhuGeneration2020a}. Furthermore, their controllability at the single qubit level enables the exploration of quantum dynamics beyond simple time evolution with a Hamiltonian, and one prominent example is dynamics involving both unitary evolution and projective measurements \cite{nahum2018hybrid, nandkishore2018hybrid, li2018hybrid, choi2019qec, gullans2019purification, andreas2019hybrid, choi2019spin}. Nevertheless, quantum devices are noisy and any states prepared are subject to decoherence.  One important question is which properties of quantum matter survive in a noisy environment, and how to understand any residual order in the mixed state?  Recently, substantial progress has been made in understanding mixed state versions of gapped quantum systems including symmetry protected topological phases and topological order \cite{deGroot2022symmetryprotected,chong,Lee2022Symmetry,Lake2022Exact,ashvin1,ashvin2,zhang2022strange,behrends2022surface}.

In this work, we study quantum critical states under decoherence, modelled by local quantum channels.  A point of inspiration is prior work \cite{Rajabpour_2016,Lin_2022_probing,Garratt_2022_measurements,swingle}  which found that measurements (with outcomes recorded) can have a significant effect on quantum critical states, in some cases boosting long range correlation.  However, verifying such effects typically requires post-selecting on monitored pure state trajectories.  In contrast, quantum channels can be viewed as environmental measurement in which the outcomes are averaged over, and we are interested in properties of the resulting mixed state, thus avoiding the need for post-selection.  


Naively one might expect that local quantum channels cannot induce new universality classes since they can be represented as finite depth unitaries acting on an enlarged Hilbert space, and such finite depth unitaries cannot change the nature of long-distance correlations.  While this is true for local, linear observables of the state, non-linear quantities like entanglement measures can be significantly affected by local channels.  

Indeed, a key finding of our work is that for entanglement measures like Renyi entropies and entanglement negativity, local quantum channels acting on quantum critical states can drive phase transitions or more technically, renormalization group (RG) flows, between different conformal fixed points labeling different quantum channels.  More specifically, we find that in computing such measures, quantum channels map to boundary conditions for multiple copies of the original conformal field theory (CFT) describing the critical state.  Under coarse-graining, such boundary conditions can flow to a variety of conformal fixed points whose classification can be much richer than that of the single copy CFT.         

Concretely, we find that the Renyi entropy of the decohered state, which quantifies the entanglement between system and environment, is extensive but has a subleading term dictated by a ``g-function'' in CFT.  This g-function characterizes the channel with respect to Renyi entropy and dictates the direction of RG flow between two channels: a channel with a larger $g$ can flow to a channel with smaller $g$ if weakly perturbed by the latter.  Moreover, near a RG fixed point, the $g$-function has a universal scaling form determined by critical exponents. One natural consequence of our formalism is that the Renyi entropy of a subsystem $A$ of the decohered state has a subleading logarithmic scaling with $|A|$ and its coefficient is given by scaling dimensions of certain boundary condition changing operators. Furthermore, Renyi negativity of the subsystem (a measure of entanglement within the mixed state) can either obey logarithmic scaling or area law based on the universal properties of the RG fixed point. In the former case, the coefficient of log scaling can either change continuously with decoherence strength or remain the same as the initial pure critical state, depending on whether the channel corresponds to a marginal or irrelevant perturbation, respectively.  We demonstrate all these features for the one dimensional transverse field Ising critical point and conclude with a discussion of experimental relevance to quantum simulators.

\emph{Quantum channels--}
We first review useful facts about quantum channels \cite{preskill1998lecture}. A quantum channel represents the most general quantum process including decoherence. Formally, it is a completely positive trace-preserving map $\rho\rightarrow\mathcal{N}(\rho)$ acting on density matrices.
For a quantum channel, one may define its dual-channel $\mathcal{N}^*$, which is a linear map on operators satisfying $\tr(\rho \mathcal{N}^{*}(O)) = \tr(\mathcal{N}(\rho) O),\ \forall O$. 

Here we consider a spatially uncorrelated noise model defined on a qubit chain. The noise can be modeled as a product of single-qubit channels $
    \mathcal{N} = \otimes_{j=1}^L \mathcal{N}_j
$
where each $\mathcal{N}_j$ is a channel acting on the $j^{\text{th}}$-qubit.

We focus on the dephasing channel $\mathcal{N}_j = \mathcal{D}^{[j]}_{p,\vec{v}}$, which represents an environment-qubit coupling with the $j$-th system qubit along the $\sigma_{\vec{v}} := v_x \sigma_x + v_y \sigma_y + v_z \sigma_z$ direction. 
The action of $\mathcal{D}^{[j]}_{p,\vec{v}}$ on the density matrix $\rho$ of a qubit is
\begin{equation}
    \mathcal{D}^{[j]}_{p,\vec{v}}(\rho) = \left(1-\frac{p}{2}\right)\rho + \frac{p}{2}\sigma^{[j]}_{\vec{v}}~\rho~\sigma^{[j]}_{\vec{v}}.
\end{equation}
The channel is self-dual, i.e., $\mathcal{D}^{[j]*}_{p,\vec{v}}=\mathcal{D}^{[j]}_{p,\vec{v}}$. The parameter $p\in[0,1]$ controls the strength of the noise: when $p=0$, it is the identity channel and leaves states unchanged; while if $p=1$, it turns any quantum state into a classical ensemble.


\emph{General formalism mapping information-theoretic quantities to quantum quench problem--}
We consider a critical ground state $|\psi\rangle$ under decoherence of the above type, resulting in density matrix $\rho=\mathcal{N}(\ket{\psi}\bra{\psi})$. We study information-theoretic quantities including $n^{\text{th}}$-Renyi entropy and entanglement negativity, and 
we map these quantities to quantum quench problems in $\text{CFT}^{\otimes 2n}$, \textit{i.e.} a $2n-$copied CFT. 

\begin{figure}[t]
    \centering
    \includegraphics[width = 0.7\linewidth]{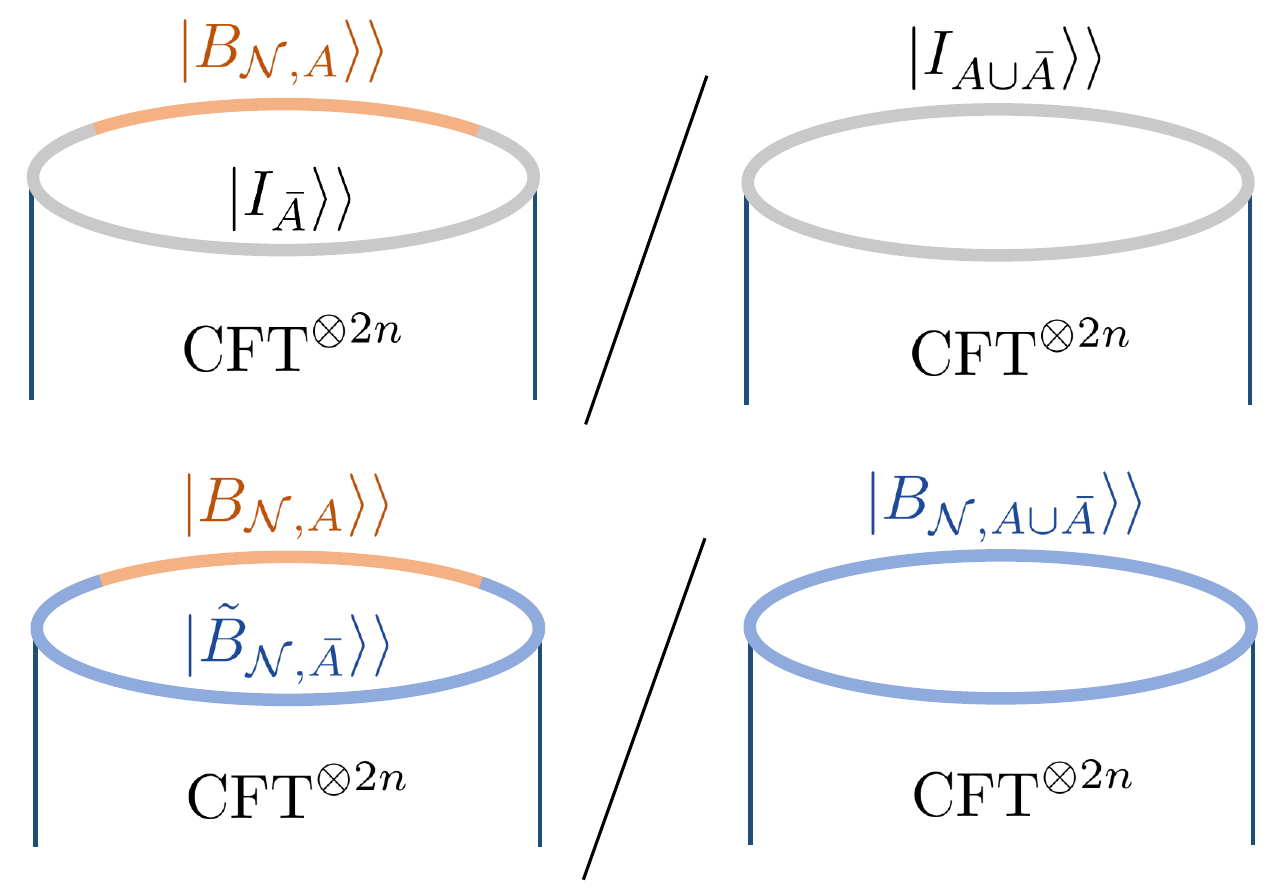}
    \caption{
   Entanglement quantities of decohered critical state involve overlaps between copies of CFT and boundary states determined by the quantum channel (see main text).    Top:  Path integral representation of partition function used to compute Renyi entropy Eq.~\eqref{eq:Zn3}. Note that the denominator is $\langle\psi^{\otimes n}|\psi^{\otimes n}\rangle=1$.
    Bottom: path integral representation of the analogous object for Renyi negativity Eq.~\eqref{eq:MN_quench}.
    }
    \label{fig:channel}
\end{figure}

We start with the Renyi entropy of a subsystem $A$: 
$   S^{(n)}_A(\rho):= \frac{1}{1-n} \log \tr (\rho_A^n),$ where $\rho_A = \tr_{\bar{A}} \rho$.
This is a measure of both quantum and classical correlation between $A$ and rest of the system $\bar{A}$ together with the environment. It reduces to the von-Neumann entropy after taking the replica limit $n\rightarrow 1$. The partition function $Z^{(n)}_A:=\tr (\rho_A^n)$ can be rewritten as:
\begin{equation}
\label{eq:Zn1}
    Z^{(n)}_A =\tr(\rho^{\otimes n} \tau_{n,A}) = \tr(\mathcal{N}(|\psi\rangle\langle \psi|)^{\otimes n} \tau_{n,A}),
\end{equation}
where $\tau_{n,A}=\prod_{j\in A}\tau_{n,j} $ forward permutes the replicas for sites in $A$, and acts as identity for sites in $\bar{A}$. Next, we use the dual channel to rewrite
\begin{equation}
\label{eq:Zn2}
     Z^{(n)}_A = \tr((|\psi\rangle\langle \psi|)^{\otimes n} B_{\mathcal{N},A}),
\end{equation}
where $B_{\mathcal{N},A} := \mathcal{N}^{*\otimes n}(\tau_{n,A})=\otimes_{j\in A} \mathcal{N}^{* \otimes n}_j(\tau_{n,j})$. 
Finally, we use the standard folding trick for defect CFT \cite{OSHIKAWA1997533} to treat each operator $O$ as a state $|O\rangle\rangle$ in the doubled Hilbert space, whose inner product is defined as $\langle \langle O_1 |O_2\rangle\rangle  = \tr(O^{\dagger}_1 O_2)$. Thus, under the folding,
\begin{equation}
\label{eq:Zn3}
    Z^{(n)}_A = \langle\langle (\psi \otimes \psi^{*})^{\otimes n}| B_{\mathcal{N},A} I_{\bar{A}}\rangle\rangle.
\end{equation}
The bra-state $|(\psi \otimes \psi^{*})^{\otimes n}\rangle\rangle$ is $2n$ copies of the critical state,  while the ket-state $| B_{\mathcal{N}}\rangle\rangle$ is a product state, since the operator $B_{\mathcal{N},A}$ factorizes among $j$. Thus, we have reformulated the Renyi entropy as a quantum quench problem from a product state $|B_{\mathcal{N}}\rangle\rangle$ to a $2n$-copied CFT, whose path integral representation on the spacetime manifold is shown in Fig.~\ref{fig:channel}(top).


\begin{figure}[t]
    \centering
    \includegraphics[width = 0.45\linewidth]{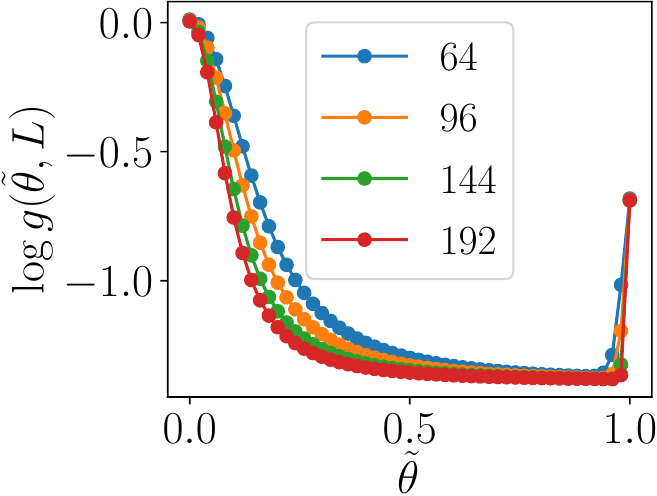}
    \hspace{0.15cm}
    \includegraphics[width = 0.45\linewidth]{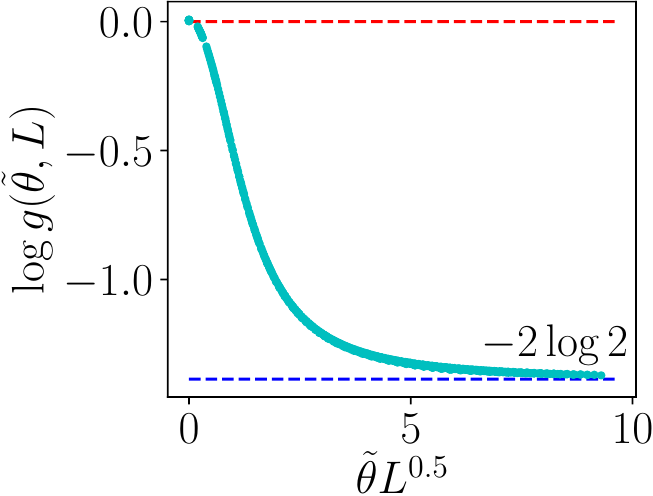}
    \caption{$g$-function (universal subleading constant of Renyi entropy) for complete dephasing channel $\mathcal{D}_{p=1,\vec{v}}$ applied to critical Ising ground state, where $\vec{v} = (\sin (\pi\tilde{\theta}/2), 0, \cos (\pi\tilde{\theta}/2) )$ in the $XZ$ plane. Left: $g(\tilde{\theta},L)$ for different $L$ (legend). Right: Data collapse for $0\leq \tilde{\theta}\leq 0.6$, indicating RG flow from $\mathcal{D}_{z}$ to $\mathcal{D}_{zx}$.}
    \label{fig:gfunction-ZX}
\end{figure}

The other quantity that we consider is $n$-th ($n$ is odd) Renyi entanglement negativity $N^{(n)}_{A}(\rho):= \frac{1}{1-n}\log \frac{\tr (\{\rho^{T_A}\}^n)}{\tr (\rho^n)}$ for a subsystem $A$, where $(\cdot)^{T_A}$ is the partial-transpose operation that swaps the bra and ket states in $\rho$ for sites within $A$. $N_A^{(n)}$ is a measure of \textit{quantum} correlation between $A$ and the rest of the system $\bar{A}$ \cite{Vidal_2002_computable,Calabrese_2012_negativity} . 
A similar derivation shows that
\begin{equation}
\label{eq:MN_quench}
N^{(n)}_{A}(\rho)=
\frac{1}{1-n}\log \frac{\langle\langle(\psi \otimes \psi^{*})^{\otimes n}|B_{\mathcal{N},A} \tilde{B}_{\mathcal{N},\bar{A}}\rangle\rangle}{\langle\langle(\psi \otimes \psi^{*})^{\otimes n}|B_{\mathcal{N},A\cup\bar{A} }\rangle\rangle}.
\end{equation}
where $\tilde{B}_{\mathcal{N},\bar{A}} = \otimes_{j\in \bar{A}} \mathcal{N}^{* \otimes n}_j(\tau^{-1}_{n,j})$.
Thus the calculation of $N_A^{(n)}$ is mapped to a quantum quench from another product initial state to the same $2n$-copied CFT, see Fig.~\ref{fig:channel}( bottom). In such quantum quench problems, it has been shown \cite{Calabrese_2006_quench,Calabrese_2007_quench,Calabrese_2016_quench,Cardy_2017_bulk} that a short-range correlated state can be described by a conformal boundary condition at long distances. We will assume that this is the case for product states $|B_{\mathcal{N}}\rangle\rangle$ and $|\tilde{B}_{\mathcal{N}}\rangle\rangle$ \footnote{See Supplemental Material for more comments on the applicability of conformal boundary conditions to quantum quench problems. }.

\emph{The $g$ function--}
We now focus on the simplest quantity
\begin{equation}
S^{(n)}(\rho)=\frac{1}{1-n}\log Z^{(n)}(L),\label{eq:Sn_from_Zn_main}
\end{equation}
where $L$ is the total system size.  This is the Renyi entropy of the whole system, or alternatively, the Renyi entanglement entropy between system and environment. 
In a boundary CFT, the partition function $Z^{(n)}(L)$ depends on UV details, but one can define a UV-independent $g$-function \cite{Friedan_2004,Casini_2016,Cuomo_2022}, 
\begin{equation}
\label{eq:g_def_main}
    \log g^{(n)}(L) = \left(1-L\frac{d}{dL}\right) \log Z^{(n)}(L).
\end{equation}
which plays a similar role for boundary RG flow as the central charge $c$-function plays for bulk RG flow. It satisfies the following two properties. First, $\log g^{(n)}(L)$ is a montonically decreasing function of $L$. Second, for conformal boundary conditions (RG fixed points), $\log g^{(n)}(L)=\log g^{(n)}$ is independent of $L$, which equals the universal Affleck-Ludwig boundary entropy \cite{Affleck:1991tk}. At these fixed points, solving the differential equation \eqref{eq:g_def_main} gives  
$\log Z^{(n)}(L) = (1-n) \alpha^{(n)}L + \log g^{(n)}$,
where $(1-n)\alpha^{(n)}$ is a non-universal integration constant. One of our main results 
\begin{equation}
\label{eq:main2}
    S^{(n)}(\rho)= \alpha^{(n)} L - \frac{1}{n-1} \log g^{(n)}_{\mathcal{N}} 
\end{equation}
then follows from Eq.~\eqref{eq:Sn_from_Zn_main}. 
Given a lattice wavefunction, we may compute $g^{(n)}_{\mathcal{N}}(L)$ numerically using Eq.~\eqref{eq:g_def_main}. We identify a channel $\mathcal{N}$ as a RG fixed point if $g^{(n)}_{\mathcal{N}}(L)$ becomes a constant $g^{(n)}_{\mathcal{N}}$ as $L$ exceeds a few lattice spacings. The identity channel $\mathcal{I}$ is always a RG fixed point with $\log g^{(n)}_{\mathcal{I}}=0$.



Let $\mathcal{N}_{p}$ be a family of quantum channels parameterized by $p$. 
We denote the $g$-function of $\mathcal{N}_p$ as $g^{(n)}_p(L)$. Let $p=p_c$ be a RG fixed point with the $g$-function $g_c$, then 
\begin{equation}
    g^{(n)}_p(L) = \bar{g}^{(n)}( |p-p_c| L^{1/\nu}),
\end{equation}
where $\nu$ is the analog of the ``correlation length exponent" in  critical phenomena. If $\nu>0$, then the fixed point $\mathcal{N}_{p_c}$ is unstable, and it flows into another fixed point with the $g$-function $g'<g_c$. In such cases, the scaling function $\bar{g}^{(n)}$ is a monotonically decreasing function.

\begin{figure*}[t]
    \centering
    \includegraphics[width = 0.3\textwidth]{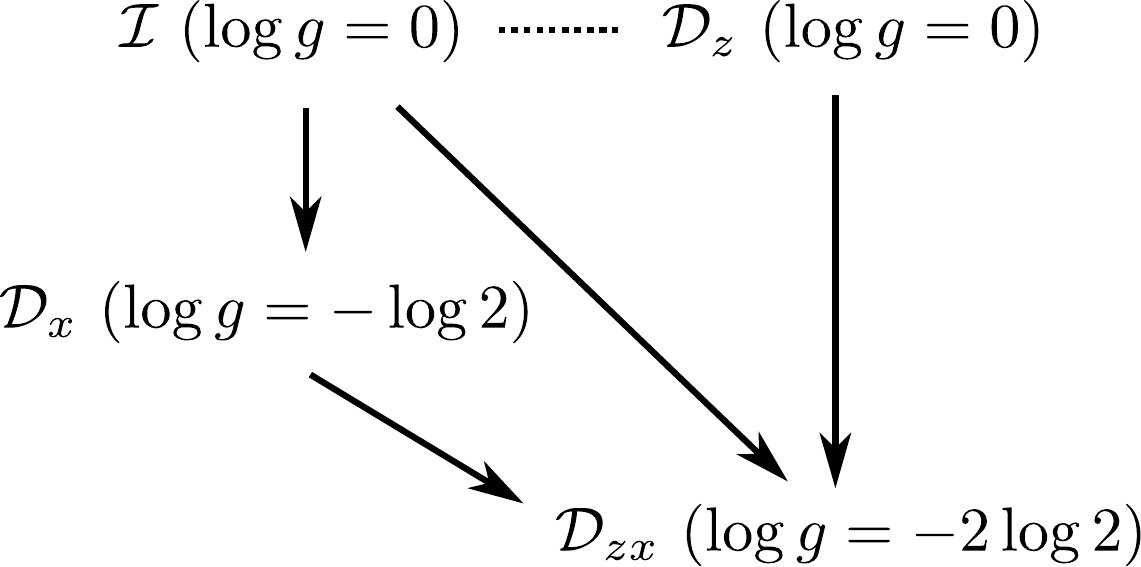}
    \hspace{0.5cm}
    \includegraphics[width = 0.5\textwidth]{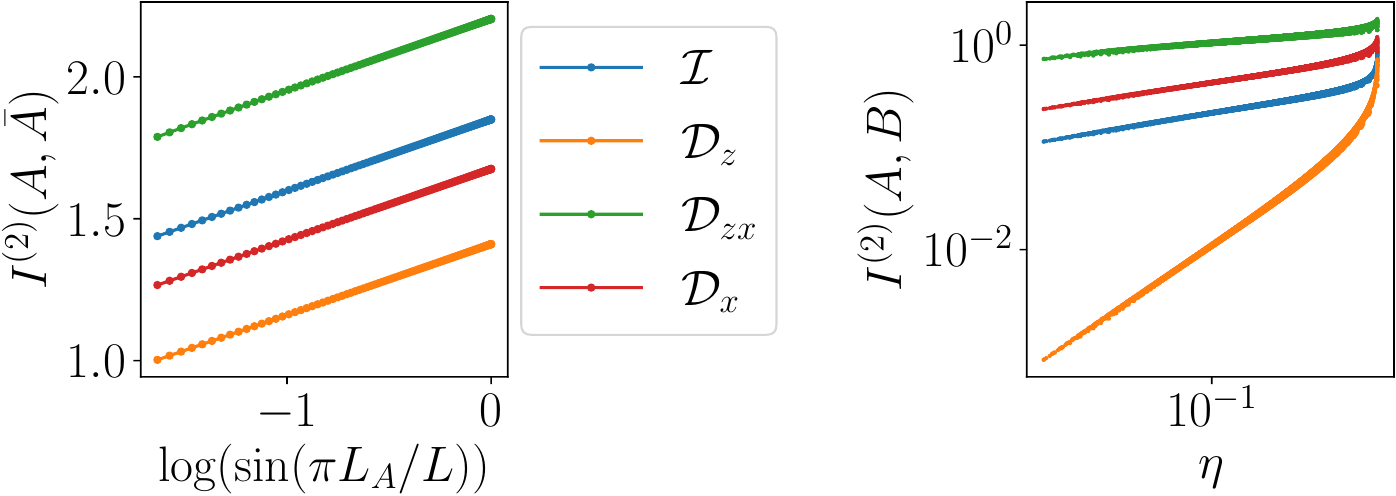}
    \caption{Fixed-point channels for Ising model and their universal properties. Left: $g$-function of the four fixed points among dephasing channels and their RG flow. The dashed line indicates a continuous set of fixed points and the solid lines indicate direction of the RG flow. Center: $I^{(2)}(A,\bar{A})$ of a single interval at the four fixed points. All four slopes are estimated to be close to $4\Delta^{(2)}_{\mathcal{I}\mathcal{N}} \approx 1/4$. Right: $I^{(2)}(A,B)$ of a two disjoint intervals at the four fixed points. The scaling dimension $\Delta^{(2)}_O$ is close to $1/4,1,1/8,1/4$ for the four fixed points $\mathcal{I},\mathcal{D}_z,\mathcal{D}_{zx},\mathcal{D}_x$, respectively. }
    \label{fig:RG-ZX}
\end{figure*}




\emph{Subsystem entropy--} Now we consider the entropy of a subsystem, whose correponding quantum quench problem is from a spatially inhomogeneous initial state, see Fig.~\ref{fig:channel}(top). 
Such initial state can be described in CFT by inserting boundary condition changing operator $\phi^{(n)}_{\mathcal{I}\mathcal{N}}$ and $\phi^{(n)}_{\mathcal{N}\mathcal{I}} = \phi^{(n)*}_{\mathcal{I}\mathcal{N}}$ at the intersections \cite{AFFLECK1997BOUNDARY,Li_2021_conformal,Calabrese_2008_domainwall}. There are $2m$ insertions if $A$ contains $m$ disjoint intervals. In particular, if $A$ is a single interval with length $L_A$, then the partition function $Z^{(n)}(L)$ is
\begin{equation}
\label{eq:Zn_domain_wall}
    \langle \phi^{(n)}_{\mathcal{I}\mathcal{N}} (0) \phi^{(n)*}_{\mathcal{I}\mathcal{N}} (L_A) \rangle = C\left( \frac{L}{\pi}\sin \frac{\pi L_A}{L}\right)^{-2\Delta^{(n)}_{\mathcal{I}\mathcal{N}}},
\end{equation}
where $\Delta^{(n)}_{\mathcal{I}\mathcal{N}}$ is the scaling dimension of $\phi^{(n)}_{\mathcal{I}\mathcal{N}}$ and $C$ is a normalization constant. Eq.~\eqref{eq:Zn_domain_wall} together with Eq.~\eqref{eq:Sn_from_Zn_main} gives another main result \footnote{The coefficient $\alpha^{(n)}$ can be understood as the line tension of creating the domain wall. It is a non-universal constant that depends on details of the lattice realization.}
\begin{equation}
\label{eq:SA_main}
    S^{(n)}(\rho_{A}) = \alpha^{(n)} L_A + \frac{2\Delta^{(n)}_{\mathcal{I}\mathcal{N}}}{n-1} \log \left( \frac{L}{\pi}\sin \left(\frac{\pi L_A}{L}\right)\right) + O(1).
\end{equation}

In order to numerically verify the formula, it is useful to consider the Renyi mutual information $I^{(n)}(A,B) = S^{(n)}(\rho_A) + S^{(n)}(\rho_B) - S^{(n)}(\rho_{AB})$.  Choosing $B=\bar{A}$ to be the complement of $A$, we obtain
\begin{equation}
\label{eq:IA_main}
    I^{(n)}(A,\bar{A}) = \frac{4\Delta^{(n)}_{\mathcal{I}\mathcal{N}}}{n-1} \log \left( \frac{L}{\pi}\sin \left(\frac{\pi L_A}{L}\right)\right) + O(1),
\end{equation}
in which the volume law pieces have cancelled each other. Similar results hold for Renyi negativity $N^{(n)}_A(\rho)$, except that the boundary condition changing operator is different.

For two disjoint intervals $A = [x_1,x_2], B = [x_3,x_4]$, we can show that $I^{(n)}(A,B)$ is only a function of the cross ratio $\eta = (X_{12}X_{34})/(X_{13}X_{24})$, where $X_{ij} = \sin (\pi |x_i-x_j|/L)$.
At small $\eta$, we may fuse the two boundary condition changing operators $\phi^{(n)}_{\mathcal{I}\mathcal{N}} \times \phi^{(n)*}_{\mathcal{I}\mathcal{N}}  =  I + O^{(n)} + \cdots $, where $O^{(n)}$ is the second lowest operator in the operator product expansion. This implies the scaling
\begin{equation}
    I^{(n)}(A,B) = \text{const.} \times \eta^{\Delta^{(n)}_{O}} ~ (\eta \ll 1).
\end{equation}
Note that $O^{(n)}$ is a boundary operator at the boundary condition $|B_\mathcal{I}\rangle\rangle$, which we may unfold to get a bulk local operator in the $n$-copied CFT. Thus, $\Delta^{(n)}_O$ must be a sum of $n$ scaling dimensions in the original CFT. 


\emph{Critical Ising model under dephasing noise--}
As an example, we study the effect of dephasing noise on the ground state of the 1d transverse field Ising model 
\begin{equation}
    H = - \sum_{i=1}^L \sigma_{x,i} \sigma_{x,i+1} - \sum_{i=1}^L \sigma_{z,i}.
\end{equation}
For the dephasing noise $\mathcal{D}_{p,\vec{v}}$, we restrict the direction to the $XZ$ plane, i.e., $\vec{v} = (\sin (\pi\tilde{\theta}/2), 0, \cos (\pi\tilde{\theta}/2) )$, where $0\leq \tilde{\theta} \leq 1$. For Renyi index $n=2$, we find four RG fixed points: (1) identity channel $\mathcal{I} = \mathcal{D}_{0,\vec{v}} $ with $\log g_{\mathcal{I}} = 0$, (2) complete dephasing in $Z$ direction $\mathcal{D}_{z}$ with $ \log g_{z} = 0$, (3) complete dephasing in $X$ direction $\mathcal{D}_{x}$ with $\log g_x=-\log 2$ and (4) complete dephasing $\mathcal{D}_{zx}$ at $\tilde{\theta}\approx 0.8$, with $\log g_{zx}= - 2\log 2$. For complete dephasing, the entropy reduces to the ``classical Shannon entropy'' of Ref.~\cite{Stephan_2009_shannon}, which considered cases (2) and (3).

We also observe the RG flow between these four fixed points, which manifests the monotonicity of the $g$-function. Near the $\mathcal{D}_{z}, \mathcal{D}_{x}$ fixed points, we can turn on a small perturbation to $\tilde{\theta}$ and observe the RG flows to the $\mathcal{D}_{zx}$ fixed point. The universal data collapses indicate the critical exponents $\nu \approx 2.0, 1.0$ for the two respective channels, see Fig.~\ref{fig:gfunction-ZX} and \cite{supp}.
Another example is the RG flow from $\mathcal{I}$ fixed point to complete dephasing $\mathcal{D}_{zx}$ and $\mathcal{D}_x$ by turning on a small $p$ \cite{supp}.
Finally, all dephasing channels $\mathcal{D}_{p,z}$ have the same $\log g=0$, which suggests they constitute a continuous family of fixed points connected by marginal deformations. These RG flows are summarized in Fig.~\ref{fig:RG-ZX}. We have also checked that the RG flows in this example are the same for higher Renyi indices such as $n=3$, although consistency between Renyi indices is not a priori true \footnote{We have observed consistency also for the dephasing in $Y$ direction, which appears to be an irrelevant perturbation for both $n=2$ and $n=3$.}. 


The mutual information between subsystems at the four RG fixed points is shown in Fig.~\ref{fig:RG-ZX}. For complementary intervals, we find that they all satisfy Eq.~\eqref{eq:IA_main} with the same $\Delta^{(2)}_{\mathcal{I}\mathcal{N}} = 1/16$. For the identity channel, the operator $\phi^{(2)}_{\mathcal{I}\mathcal{N}}$ is the branch-point twist operator, with the scaling dimension $c/8 = 1/16$. The fact that the scaling dimensions coincide for dephasing channel and identity channel was also observed numerically in Ref.~\cite{Alcaraz_2013_universal}. In order to distinguish the channels, we compute the Renyi mutual information $I^{(n)}(A,B)$ of two disjoint intervals. We see that the scaling dimensions of $\Delta^{(2)}_O$ vary among the four fixed points. These dimensions can be understood analytically in terms of local operators in the two-copied Ising model \footnote{See \cite{supp} for details, which contain references \cite{Shapourian_2017_negativity,Liu_2022_vertex,Zou_2020_conformal,Shapourian2019PRA,Shapourian2019scipost,Murciano2022,Cardy_2017_bulk,zhuGeneration2020a,anand2022holographic,Huang_2020_predicting,Elben_2020_mixed,Li2022Entanglement,10.21468/SciPostPhys.6.3.029}.}.

\begin{figure}
    \centering
    \includegraphics[width = 0.9 \linewidth]{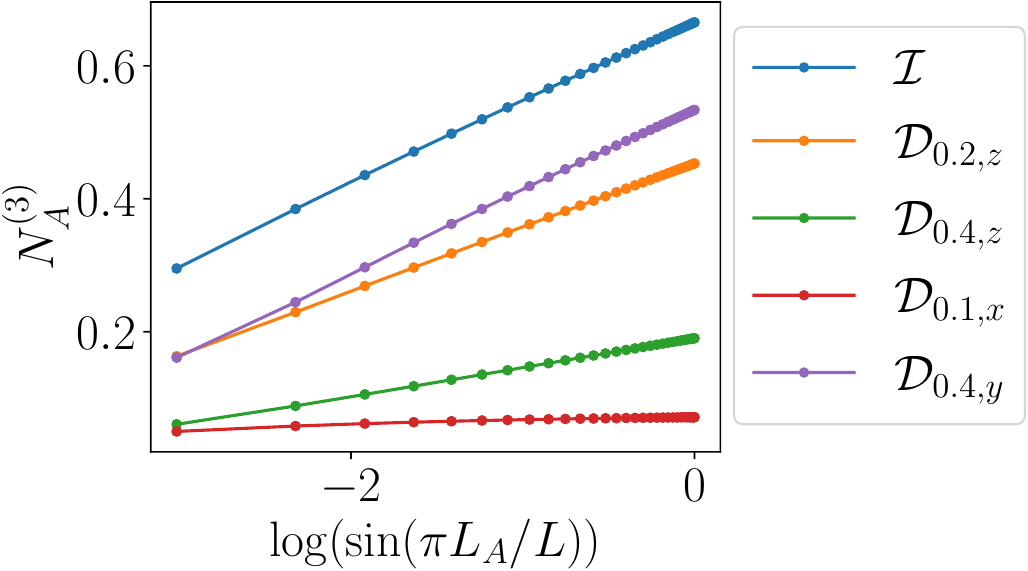}
    \caption{Renyi negativity $N^{(3)}_A$ of a subsystem $A$ for the ground state of critical Ising model ($L=64$) under various choices of dephasing noise. Dephasing in $x,y,z$ directions are respectively relevant, irrelevant, and marginal perturbations, implying that the negativity scaling is respectively area law or log scaling with either the same coefficient as the initial pure critical state or continuously changing with strength $p$.}
    \label{fig:NT}
\end{figure}

Finally, we study the Renyi negativity $N^{(3)}_A$ of a subsystem $A$ under dephasing noise, see Fig.~\ref{fig:NT}. For $Z$-dephasing, we find a logarithmic scaling with continuously changing coefficients, which is an indication of a continuous set of RG fixed points. For $X$-dephasing, even with a small strength $p$, the Renyi negativity obeys an area law, in agreement with the previous observation that the channel flows to complete dephasing. We also observe that a weak $Y$-dephasing is an irrelevant perturbation, indicated both by the $g$-function and the scaling of $I^{(2)}(A,\bar{A})$ and $N^{(3)}(A,\bar{A})$. Indeed, we find in Fig.~\ref{fig:NT} that for $Y$-dephasing with small $p$, the coefficient of the log scaling of negativity does not change with respect to the initial pure critical state, i.e. the one with $\mathcal{I}$ channel.


\emph{Discussion--}
In this work we have established a notion of RG flow between quantum channels acting on a critical wavefunction, and we have demonstrated its consequences on the entanglement structure of the resulting mixed states. We have shown that a $g$-function gives the volume-independent part of the  Renyi entropy at the RG fixed points and established its monotonicity under the RG flow. We have illustrated all this in the example of the transverse field Ising model under dephasing noise, although generalization to other noise models such as depolarizing noise is straightforward and also yields interesting phenomena (see \cite{supp} for details).

Our notion of RG flow can be naturally applied to measurement-induced quantities of a critical wavefunction, such as those considered in Refs. \cite{Lin_2022_probing,Garratt_2022_measurements}. Another direction is to classify the conformal boundary conditions for the fixed-point quantum channels in the CFT and take the replica limit, along the lines of \cite{Quella_2007}. 
Regarding the holographic correspondence between a CFT on the boundary of a spacetime and a gravitational theory in the bulk, it would be interesting to explore the holographic dual of quantum channels acting on the boundary; many recent works have considered coupling the boundary CFT to a bath, and  perhaps our formalism could be related to replica wormholes \cite{holo1, holo2, holo3}.  In another direction, while this work only considered one layer of local quantum channels, it may also be interesting to explore the entanglement dynamics of random local channels \cite{Li2022Entanglement} on quantum critical (and other long-range entangled) states.

Finally, we discuss the  experimental relevance of our work.  Quantum critical states have already been realized on programmable quantum simulators (see e.g. \cite{zhuGeneration2020a,anand2022holographic} for the realization of a critical transverse-field Ising ground state), and such platforms may be naturally subject to biased noise in which one type of channel (e.g. dephasing in a particular direction) dominates (see \cite{bias1,bias2,bias3} for details on different hardwares).  Alternatively, the noise channels can also be realized by local unitary gates acting on system and ancillae.  The entanglement properties (Renyi entropy and negativity) of the resulting mixed states can be probed using recent techniques of shadow tomography \cite{Huang_2020_predicting,Elben_2020_mixed} which use randomized measurements to estimate moments of a (partially transposed) density matrix. Our predictions can be realized on near-term quantum simulators, and we discuss the experimental setup in detail in \cite{supp}. 

\emph{Acknowledgements--}
We thank  Runze Chi, Soonwon Choi, Paolo Glorioso, Vedika Khemani, Yaodong Li, Yuhan Liu, Peter Lu, Alex May,  Xiao-Liang Qi, Guifre Vidal, Jinzhao Wang and Fei Yan for helpful discussions. Y.Z. particularly thanks Ehud Altman, Samuel Garatt and Zack Weinstein for explanation of their work Ref.~\cite{Garratt_2022_measurements}. We thank Jacob Lin and Weicheng Ye for collaboration on a related project. S.S. and T.H. are supported by Perimeter Institute and NSERC.  Research at Perimeter Institute is supported in part by the Government of Canada through the Department of Innovation, Science and Economic Development Canada and by the Province of Ontario through the Ministry of Colleges and Universities. Y.~Z. is supported by the Q-FARM fellowship at Stanford University.

\emph{Note added:} While completing this manuscript, we noticed a related independent work \cite{critical}. 
\bibliography{refs}

\newpage

\onecolumngrid
\appendix

\section*{Ising model with dephasing and depolarizing noise }
In this appendix, We consider dephasing and depoloarizing channels and their flow to the fixed points. The dephasing channel on site $j$ is defined by
\begin{equation}
    \mathcal{D}^{[j]}_{p,\vec{v}}(\rho) = \left(1-\frac{p}{2}\right)\rho + \frac{p}{2}\sigma^{[j]}_{\vec{v}}~\rho~\sigma^{[j]}_{\vec{v}}.
\end{equation}
The channel on the spin chain is the tensor product $\mathcal{N}_{p,\vec{v}}= \otimes_{j=1}^L \mathcal{D}^{[j]}_{p,\vec{v}}(\rho)$. We fix $\vec{v}$ and continuously change $p$ away from the fixed point $p_c=0$. We compute the $g$-function for the second Renyi entropy. As shown in the main text, the $g$-function has a universal form
\begin{equation}
    g^{(2)}(p,L) = g^{(2)}(pL^{1/\nu}),
\end{equation}
where $\nu$ depends on the direction $\vec{v}$. If $\nu>0$, then the dephasing channel is a relevant perturbation, and the data collapse indicates flow from $\mathcal{I}$ to another IR fixed-point channel. If $\nu<0$, then the dephasing channel is an irrelevant perturbation, and the $\mathcal{I}$ channel is the IR fixed point.
\subsection{Dephasing in the $ZX$ plane}
\begin{figure}[htbp]
    \centering
    \includegraphics[width = 0.32\linewidth]{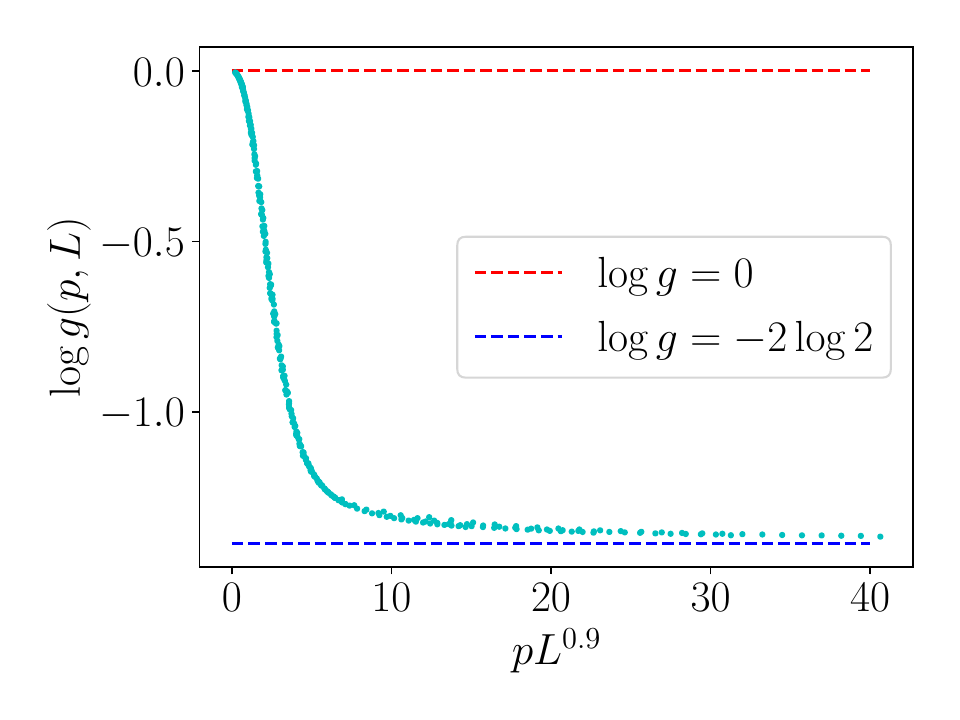}
    \includegraphics[width = 0.32\linewidth]{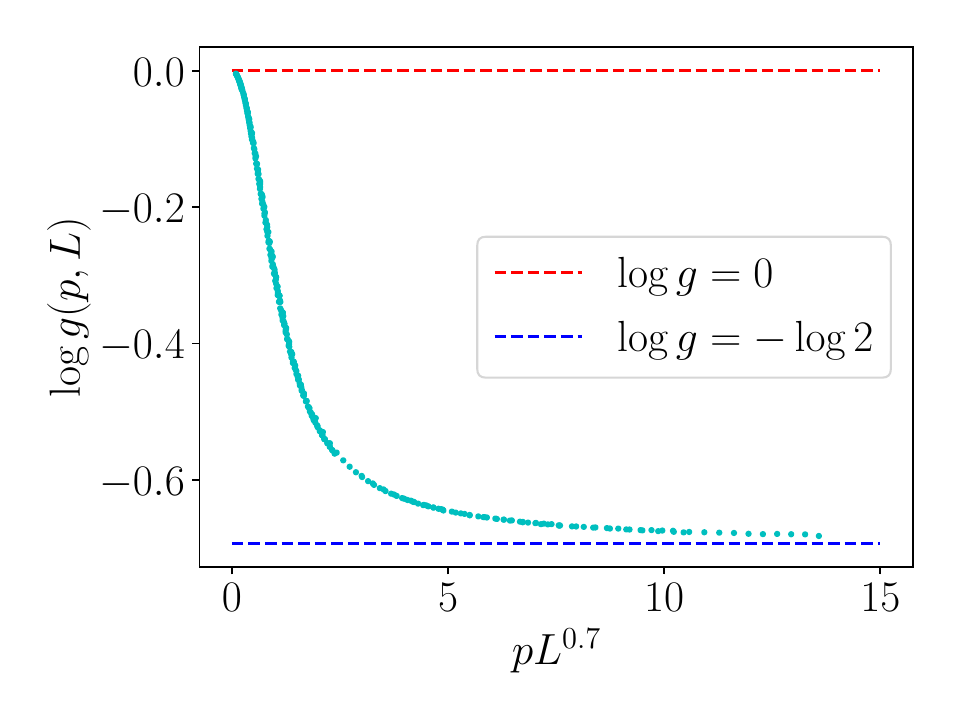}
    \includegraphics[width = 0.32\linewidth]{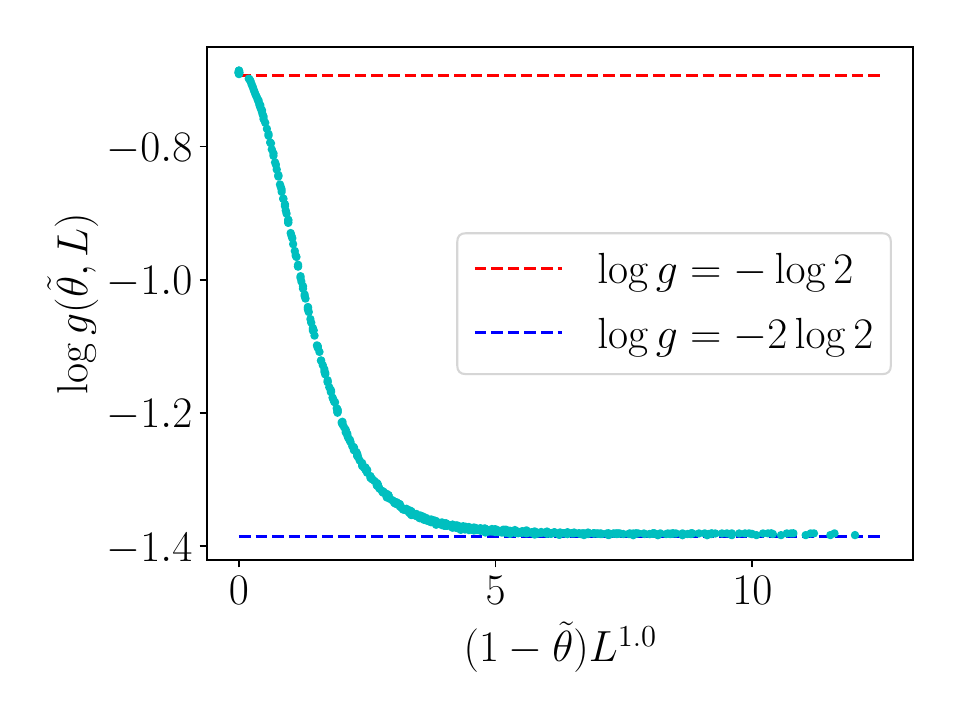}
    \caption{$g$-function of dephasing channel $\mathcal{D}_{p,\vec{v}}$. Left: Fixed $\tilde{\theta} = 0.6$ in the $ZX$ direction and varying $p$ with $0\leq p \leq 0.3$, indicating RG flow from $\mathcal{I}$ to $\mathcal{D}_{zx}$.  Center: Fixed $\tilde{\theta}=1$ and varying $p$ with $0\leq p \leq 0.3$,  indicating RG flow from $\mathcal{I}$ to $\mathcal{D}_{x}$. Right: Fixed $p=1$ and varying $\tilde{\theta}$ with $0.95\leq \tilde{\theta}\leq 1$, indicating RG flow from $\mathcal{D}_{x}$ to $\mathcal{D}_{zx}$.}
    \label{fig:gfunction-ZX-incomplete}
\end{figure}

By fixing $\tilde{\theta}$ and tuning the strength $p$, we observe the flow from $\mathcal{I}$ to complete dephasing $\mathcal{D}_{zx}$ and $\mathcal{D}_{x}$, where the critical exponents are $1/\nu \approx 0.9$ and $1/\nu \approx 0.7$, respectively. The data collapse is shown in Fig.~\ref{fig:gfunction-ZX-incomplete}. By fixing $p=1$ and tuning $\tilde{\theta}$, we also observe RG flow for complete dephasing channels from $\mathcal{D}_{z}$ and $\mathcal{D}_{x}$ to $\mathcal{D}_{zx}$. The data collapse for the $\mathcal{D}_{z}\rightarrow \mathcal{D}_{zx}$ is shown in Fig. 2(b) in the main text, indicating $1/\nu\approx 0.5$. The data collapse for the $\mathcal{D}_{x}\rightarrow \mathcal{D}_{zx}$ is shown in Fig.~\ref{fig:gfunction-ZX-incomplete}, indicating $1/\nu\approx 1.0$.

\subsection{Dephasing in $Y$ direction}
\begin{figure}[htbp]
    \centering
    \includegraphics[width = 0.4\linewidth]{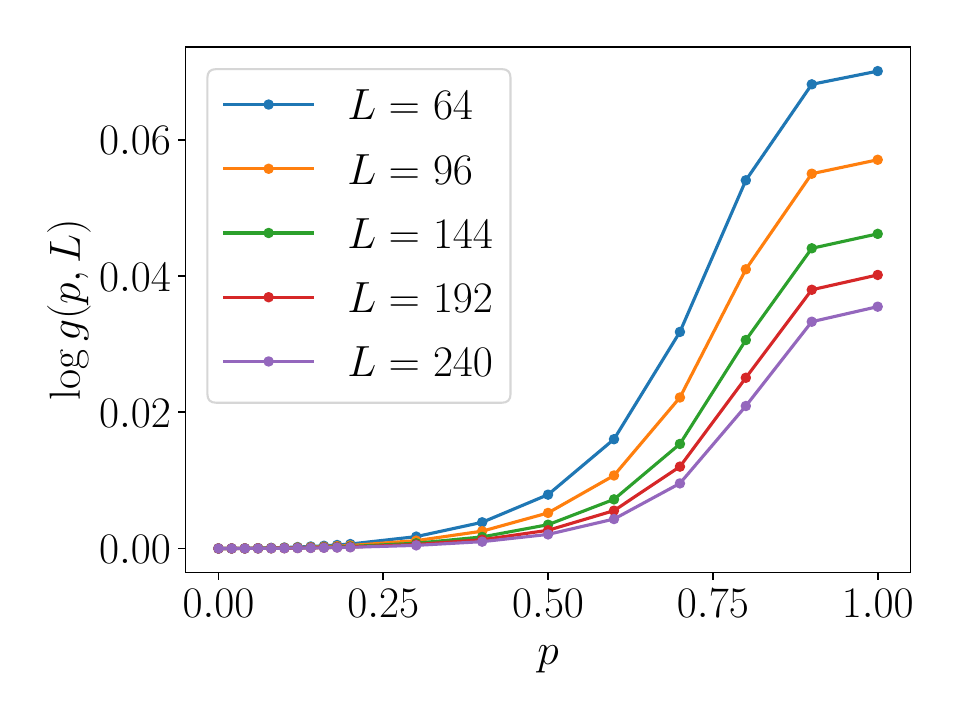}
    \includegraphics[width = 0.4\linewidth]{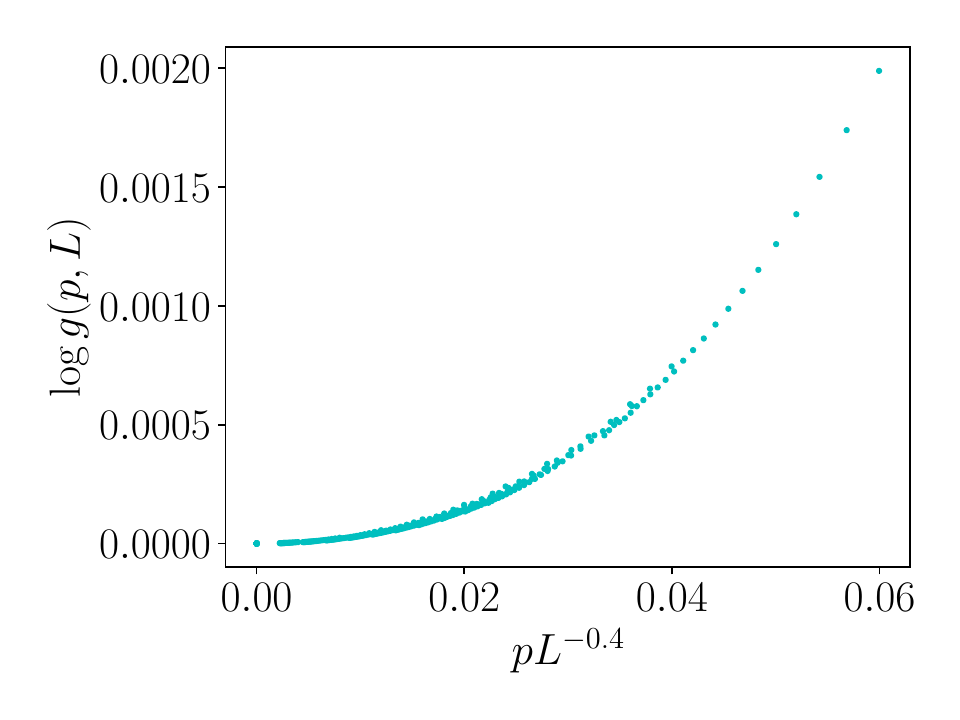}
    \caption{$g$-function of dephasing channel $\mathcal{D}_{p,\vec{y}}$. Left: $g$-function for different system sizes. Right: Data collapse near for $0\leq p \leq 0.3$, indicating RG flow to the $\mathcal{I}$ channel by an irrelevant perturbation.}
    \label{fig:gfunction-Y-incomplete}
\end{figure}

Next, we study dephasing in $Y$ direction. The $g$-function is shown in Fig.~\ref{fig:gfunction-Y-incomplete}. The data collapse shows that the dephasing is an irrelevant perturbation, where the critical exponent is $1/\nu \approx -0.4$. The nature of irrelevant perturbation suggests that a finite dephasing channel $\mathcal{D}_{p,\vec{y}}$ flows to the $\mathcal{I}$ if $p\leq p_c$, where $p_c$ is a finite threshold. Our numerical results of subsystem entropy (shown in the next section) suggests that $p_c\approx 0.6$. 

\subsection{Depolarizing noise}
\begin{figure*}[t]
    \centering
    \includegraphics[width = 0.3\linewidth]{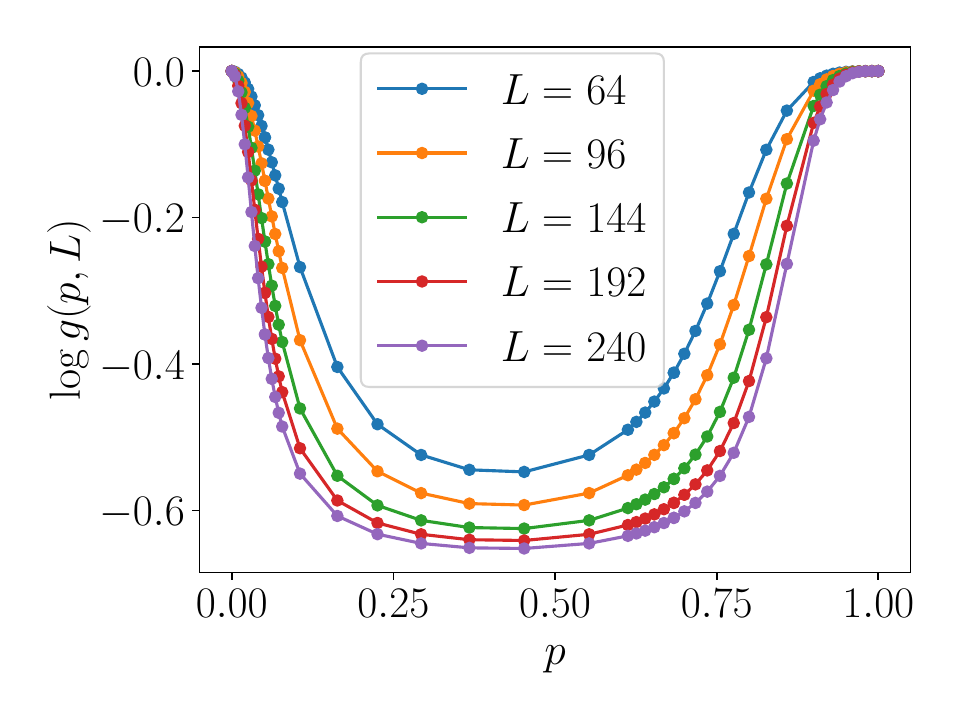}
    \includegraphics[width = 0.3\linewidth]{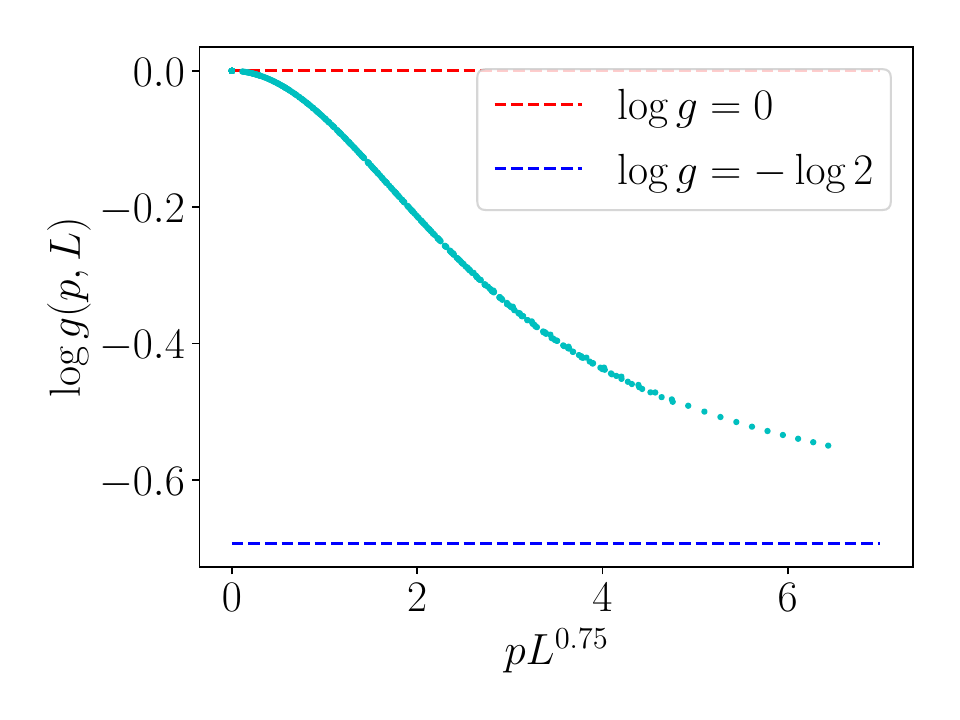}
    \includegraphics[width = 0.3\linewidth]{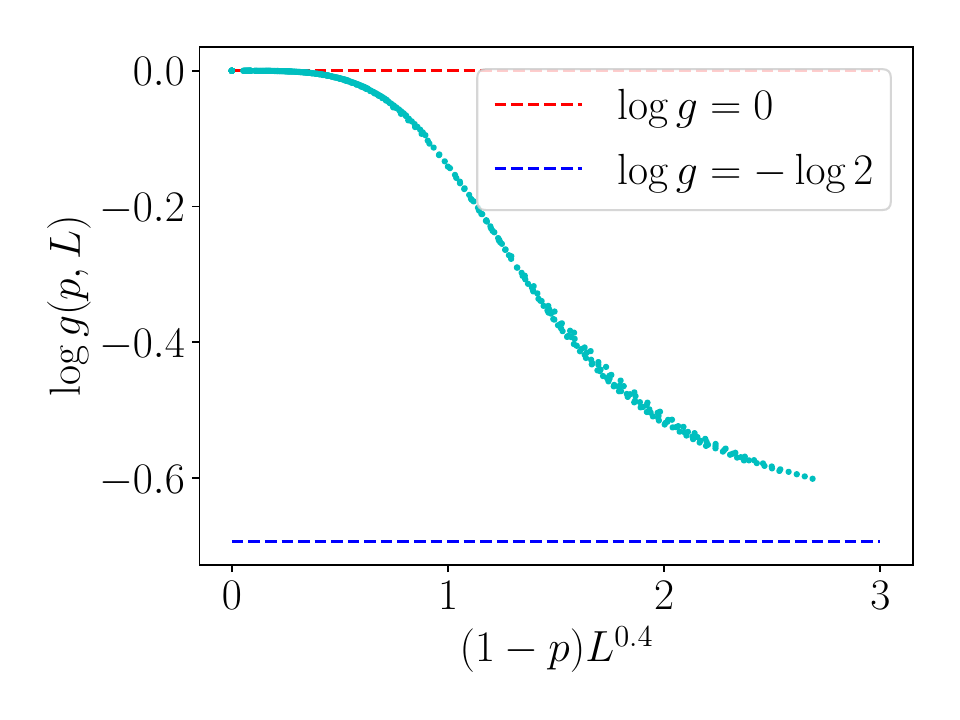}
    \caption{$g$-function of depolarization channel $\Delta_p$. Left: $g(p,L)$ for different $L$. Center: Data collapse for $0\leq p \leq 0.15$, indicating RG flow from $\mathcal{I}$ to $\Delta$. Right: Data collapse for $0.7\leq p \leq 1$, indicating RG flow from $\Delta_1$ to $\Delta$.}
    \label{fig:gfunction-depo}
\end{figure*}

Finally, we consider a different class of noise channels, the depolarizing noise, whose action on the $j$-th qubit is
\begin{equation}
    \Delta^{[j]}_{p}(\rho) = \left(1-p\right)\rho + \frac{p}{2} I.
\end{equation}
At $p=0$, the channel is the $\mathcal{I}$ channel, which is a fixed point with $\log g = 0$. At $p=1$, the channel turns any state into a maximally mixed state, thus it is also a fixed point with $\log g =0$. We denote the latter fixed point by $\Delta_1$. Surprisingly, we find another fixed point at $p\approx 0.4$, which has $\log g^{(2)} \approx -\log 2$, see Fig.~\ref{fig:gfunction-depo}. We denote this fixed point as $\Delta$. By turning on a small perturbation to $p$ at either the $\mathcal{I}$ ($p=0$) or $\Delta_1$ ($p=1$) fixed point, the channel flows to the new fixed point $\Delta$. The data collapse shows that the critical exponent is $1/\nu\approx 0.75$ and $1/\nu\approx 0.4$, respectively.

We suspect that the finite-depolarization fixed point $\Delta$ coincides with the complete dephasing $\mathcal{D}_x$ fixed point for the ground state of transverse-field Ising model. They both have $\log g^{(2)} \approx -\log 2$, and the scaling dimension of the boundary condition changing operator $\phi^{(2)}_{\mathcal{I}\mathcal{N}}$ are both approximately $1/16$. Furthermore, the RG flow from $\mathcal{I}$ to both fixed point has apprximately equal critical exponent $\nu$, and the subsystem entropy both shows a transient scaling  with close transient scaling dimensions (see the section below). We leave it as future work to determine whether $\Delta$ is the same fixed point as $\mathcal{D}_x$ from the boundary CFT perspective.

\section*{Transient behavior of subsystem entropy}
We have shown that at the fixed points the subsystem entropy satisfies a simple formula
\begin{equation}
\label{eq:IA_app}
    I^{(n)}(A,\bar{A}) = \frac{4\Delta^{(n)}_{\mathcal{I}\mathcal{N}}}{n-1} \log \left( \frac{L}{\pi}\sin \left(\frac{\pi L_A}{L}\right)\right) + O(1).
\end{equation}
Away from the fixed points, the formula is expected to hold in the limit of $L_A\rightarrow\infty,L\rightarrow\infty$ while keeping $L_A/L$ a $O(1)$ constant. However, when the channel is in the middle of a RG flow between fixed points, interesting transient behaviors occur at finite sizes. We show numerically that Eq.~\eqref{eq:IA_app} still approximately holds, but with scaling dimensions $\Delta^{(n)}_{\mathcal{I}\mathcal{N}}$ drifting with the system sizes. We study two examples of RG flow in more detail below.
\subsection{RG flow from $\mathcal{D}_x$ to $\mathcal{D}_{zx}$}
\begin{figure}[tbp]
    \centering
    \includegraphics[width = 0.99 \linewidth]{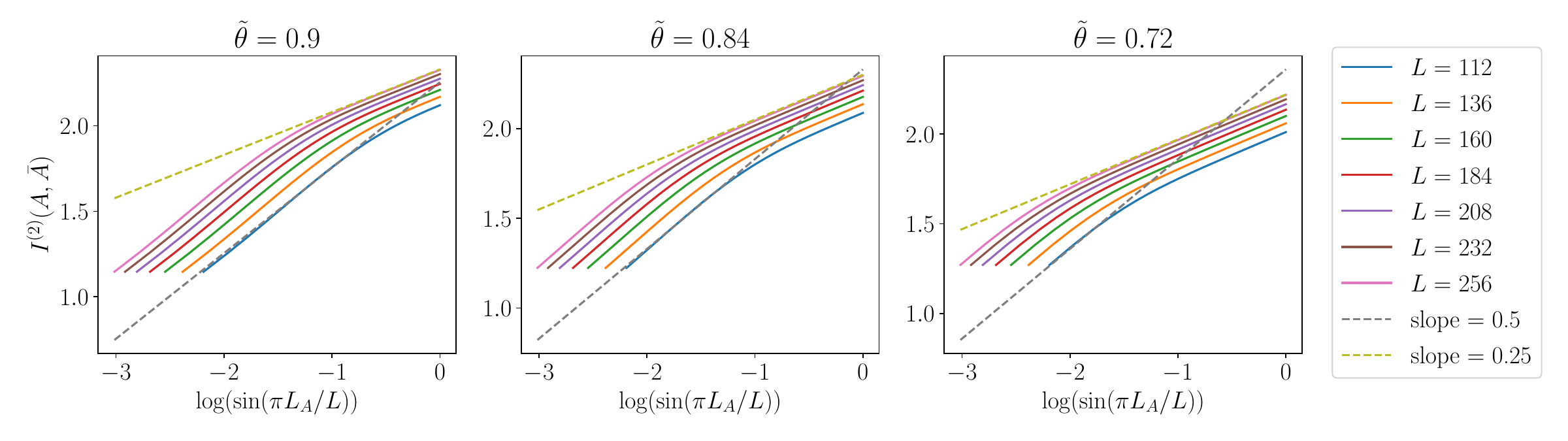}
    \caption{Renyi mutual information $I^{(2)}(A,\bar{A})$ for the complete dephasing channel acting on the ground state of Ising model.}
    \label{fig:subent_Zx}
\end{figure}

As a first example, we study the complete dephasing channel $\mathcal{D}_{p,\vec{v}}$ with $p=1$ and $\vec{v} = (\sin (\pi\tilde{\theta}/2), 0, \cos (\pi\tilde{\theta}/2) )$ acting on the ground state of the critical Ising model. We choose $\tilde{\theta}$ close to $1$ to observe a RG flow from the UV fixed point $\mathcal{D}_x$ to IR fixed point $\mathcal{D}_{zx}$. We plot the Renyi mutual information $I^{(2)}(A,\bar{A})$ in Fig.~\ref{fig:subent_Zx}. 

We see that at large $L_A$ and $L$ the subsystem entropy agrees with the IR fixed point with $4\Delta^{(2)}_{\mathcal{I}\mathcal{N}} = 1/4$. At smaller sizes we see a transient scaling with $4\Delta^{(2)}_{\mathcal{I}\mathcal{N}} = 1/2$. The crossover between the two scaling can be tuned by varying $\tilde{\theta}$. The length scales where we observe the transient scaling become broader if we tune $\tilde{\theta}$ closer to the UV fixed point $\tilde{\theta}=1$, as the RG flow becomes longer. We conjecture that the operator with $4\Delta^{(2)}_{\mathcal{I}\mathcal{N}} = 1/2$ is the subleading boundary condition changing operator from $|B_\mathcal{I}\rangle\rangle$ to $|B_\mathcal{N}\rangle\rangle$ at the IR fixed point, where $\mathcal{N} = \mathcal{D}_{zx}$. We leave it for future work to identify these operators in the replicated CFT.
\subsection{RG flow from $\mathcal{I}$ to $\Delta$}
\begin{figure}[tbp]
    \centering
    \includegraphics[width = 0.99 \linewidth]{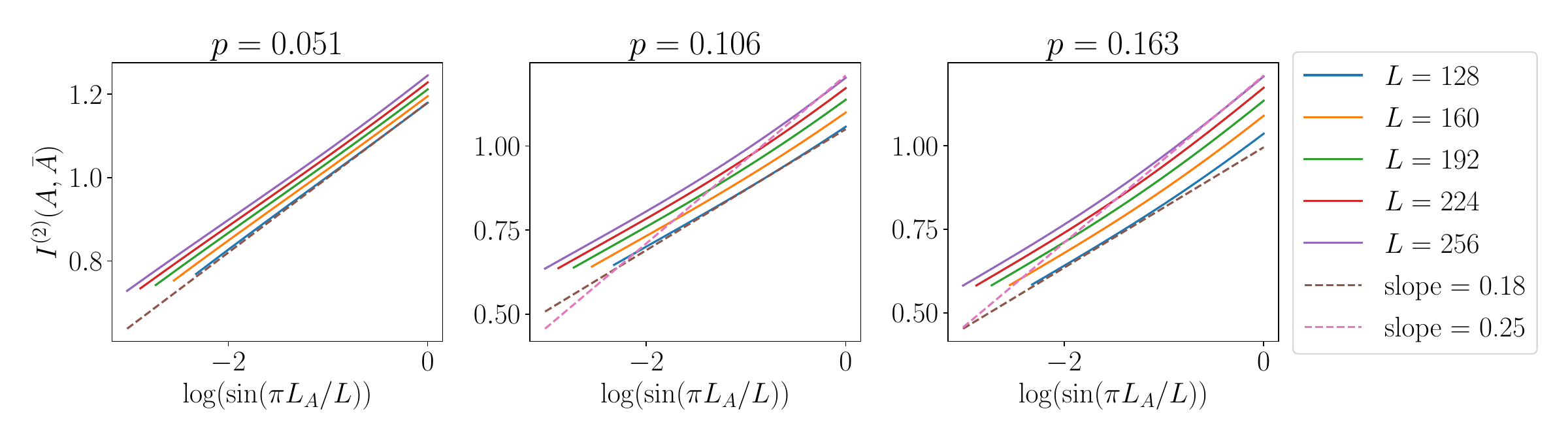}
    \caption{Renyi mutual information $I^{(2)}(A,\bar{A})$ for the depolarization channel acting on the ground state of Ising model.}
    \label{fig:subent_depo}
\end{figure}

Next, we consider the RG flow from identity channel $\mathcal{I}$ to finite depolarization $\Delta$ fixed point. We consider depolarization channel $\Delta_p$ with $p=0.1,0.2,0.3$ and plot the Renyi mutual information $I(A,\bar{A})$ in Fig.~\ref{fig:subent_depo}. Again, we see that there is a transient behavior at small sizes near the UV fixed point. At $p=0.051$, Eq.~\eqref{eq:IA_app} holds with $4\Delta^{(2)}_{I\mathcal{N}}\approx 0.18$ for all sizes $L\leq 256$. At larger $p$ and larger sizes, the IR scaling with $4\Delta^{(2)}_{I\mathcal{N}}\approx 1/4$ starts to emerge. The appearance of drifting scaling dimension at small sizes is similar to the previous example. 

We note that the same transient behavior is observed for finite dephasing channel $\mathcal{D}_{p,x}$. At small $p$ and small system sizes, e.g., $p = 0.051$ and $L\leq 256$, we also observe that Eq.~\eqref{eq:IA_app} holds with $4\Delta^{(2)}_{I\mathcal{N}}\approx 0.18$. This provide additional evidence that $\Delta$ and $\mathcal{D}_x$ may be the same fixed point, although a more detailed field-theoretic computation must be done in order to show it. 

\subsection{Dephasing in $Y$ direction}
\begin{figure}[htbp]
    \centering
    \includegraphics[width = 0.4\linewidth]{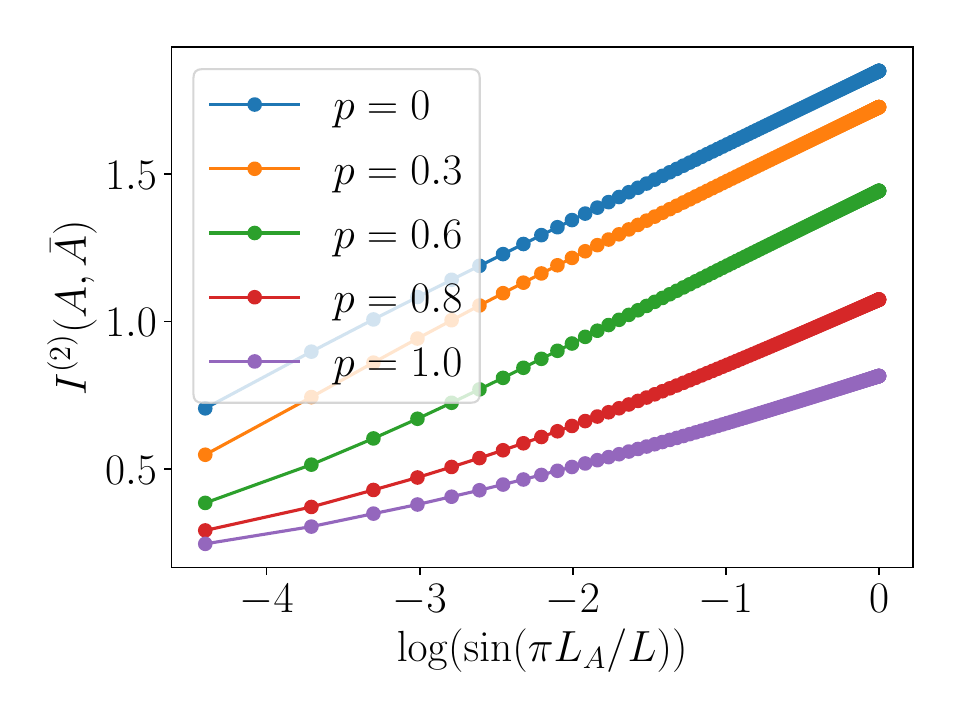}
    \includegraphics[width = 0.4\linewidth]{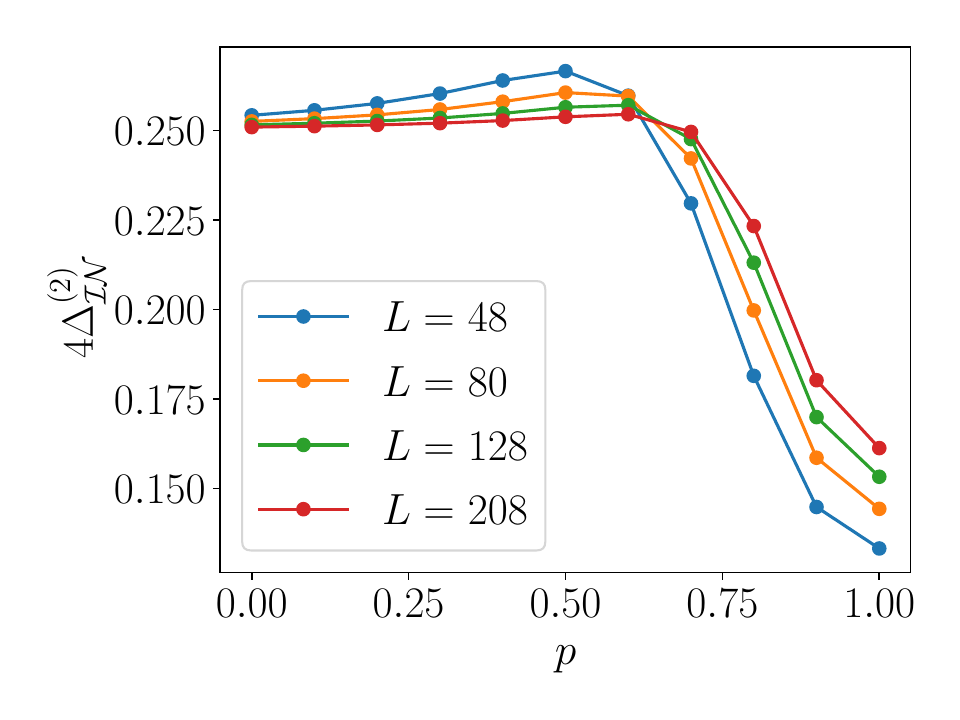}
    \caption{Renyi mutual information $I^{(2)}(A,\bar{A})$ for the ground state of Ising model acted with dephasing in $Y$ direction. Left: We fix total system size $L=256$ and change the subsystem size $L_A$ and the dephasing strength $p$. Right: We fit  and plot the scaling dimension $4\Delta^{(2)}_{I\mathcal{N}}$ in Eq.~\eqref{eq:IA_app} using data of different system sizes.}
    \label{fig:subent-Y-incomplete}
\end{figure}

We consider the Renyi mutual information $I^{(2)}(A,\bar{A})$ of the critical state dephased in the $Y$ direction, see Fig.~\ref{fig:subent-Y-incomplete}. For $p\leq 0.6$, the coefficient in front of the logarithm seems to be constant for different $p$'s, in agreement with an irrelevant perturbation. For $p>0.6$, the scaling dimension of the boundary condition changing operator $\Delta^{(2)}_{I\mathcal{N}}$ depends strongly on the system size, indicating that the channel is far away from a RG fixed point. Due to strong finite-size effect, it is currently unclear what fixed points they flow into and what universal properties these fixed points have.

\subsection{Renyi negativity}
\begin{figure}[htbp]
    \centering
    \includegraphics[width = 0.32\linewidth]{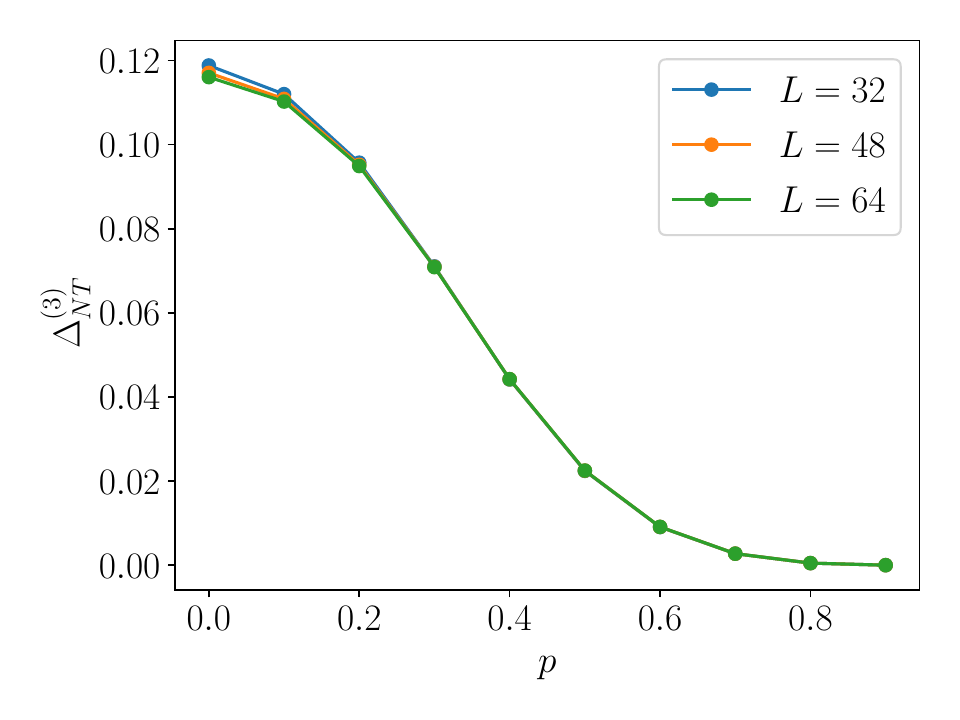}
    \includegraphics[width = 0.32\linewidth]{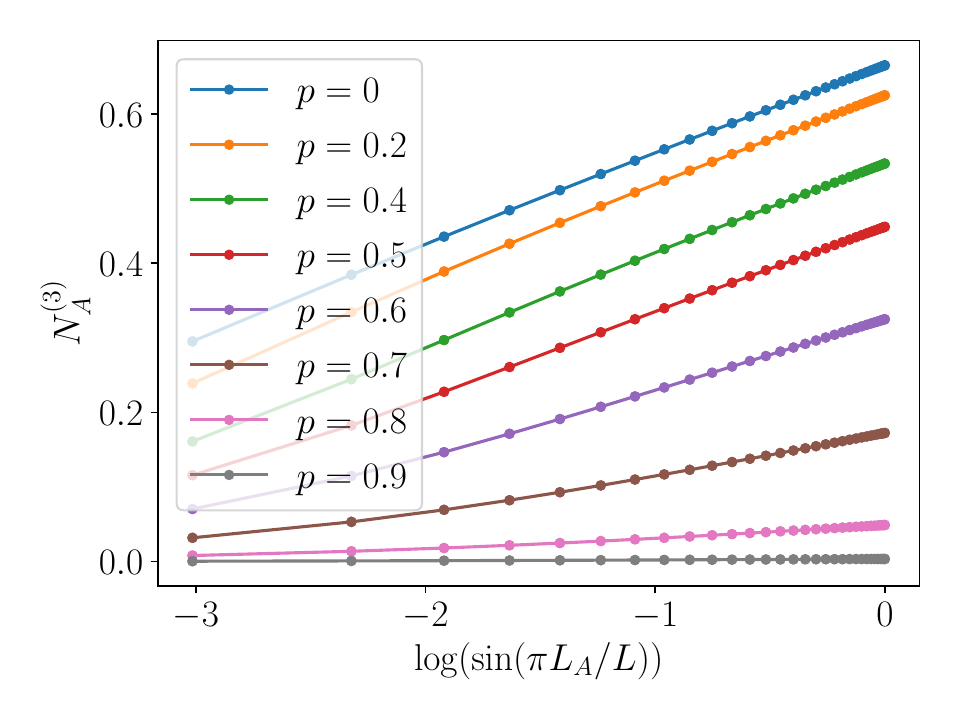}
    \includegraphics[width = 0.32\linewidth]{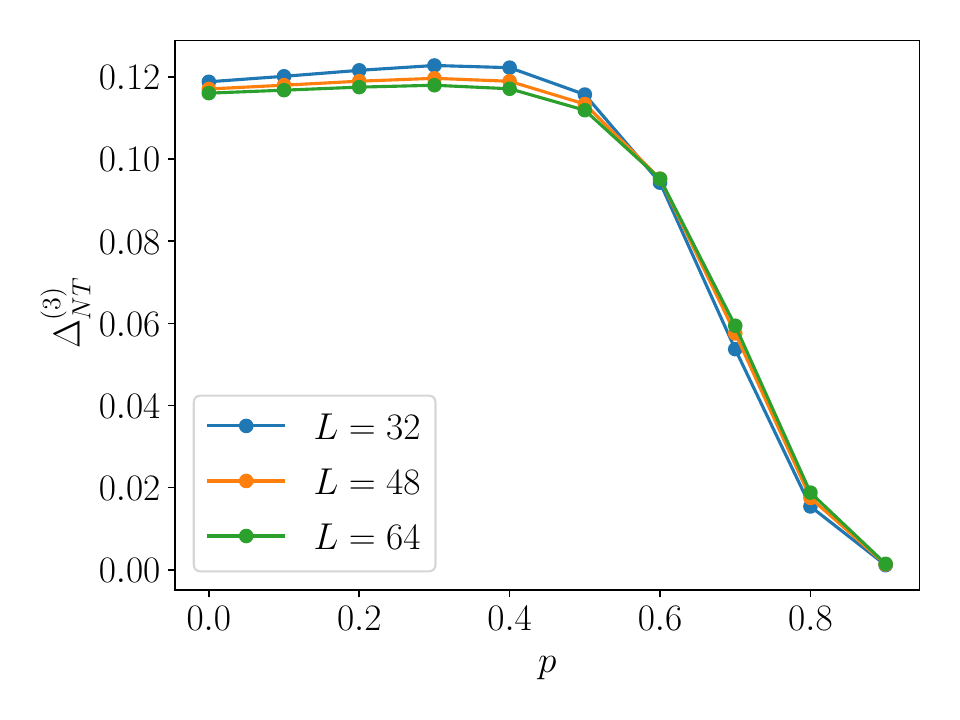}
    \caption{Renyi negativity $N^{(3)}(A,\bar{A})$ for the ground state of Ising model acted with dephasing channels. Left: The scaling dimension $\Delta^{(3)}_{NT}$ for dephasing in $Z$ direction. Center: Dephasing in $Y$ direction at $L=64$. Right: The scaling dimension $\Delta^{(3)}_{NT}$ for dephasing in $Y$ direction.}
    \label{fig:NT-Y-incomplete}
\end{figure}

Finally, we consider the Renyi negativity $N^{(3)}(A,\bar{A})$ of the critical state under dephasing noise $\mathcal{D}_{p,\vec{v}}$. As discussed in the main text, the Renyi negativity becomes area law if $\vec{v}$ is in the $ZX$ plane, except when $\vec{v} = \vec{z}$. This is because dephasing channels in these directions are relevant and flow to complete dephasing. For dephasing in the $Z$ direction, we expect that
\begin{equation}
\label{eq:NT3Y}
    N^{(3)}(A,\bar{A}) = \Delta^{(3)}_{NT} \log \left( \frac{L}{\pi}\sin \left(\frac{\pi L_A}{L}\right)\right) + O(1),
\end{equation}
where $\Delta^{(3)}_{NT}$ is the scaling dimension of the boundary condition changing operator from $|B^{(3)}_{\mathcal{N}}\rangle\rangle$ to $|\tilde{B}^{(3)}_{\mathcal{N}}\rangle\rangle$. Since dephasing in $Z$ direction is a marginal perturbation, we expect that the scaling dimension $\Delta^{(3)}_{NT}$ changes continuously with the channel strength $p$. This is indeed the case, as shown in Fig.~\ref{fig:NT-Y-incomplete}.

For dephasing in $Y$ direction, since it is an irrelevant perturbation, we expect that $\Delta^{(3)}_{NT} = \Delta^{(3)}_T = 2c/9$ remains a constant as long as $p\leq p_c$, where $\Delta^{(3)}_T$ is the scaling dimension of the branch-point twist operator at Renyi index $n=3$. For $p>p_c$, we observe that Eq.~\eqref{eq:NT3Y} still holds, but the scaling dimension $\Delta^{(3)}_{NT}$ appears to be continuously decreasing with $p$, see the center and right panels of Fig.~\ref{fig:NT-Y-incomplete}.

\section*{Free fermion with amplitude damping noise}
In this appendix, we consider the free fermion CFT and a special instance of Gaussian channel. The covariance matrix formalism makes it easy to access all Renyi entropy and negativity. In particular, we can directly compute the Von Neumann entanglement entropy and negativity, which is hard for a tensor network state.

We consider the free Majorana fermion with Neveu-Schwarz boundary condition
\begin{equation}
    H = -2i \sum_{j=1}^{2L} \psi_j \psi_{j+1},
\end{equation}
where $\psi_j$ is the Majorana operator that satisfies $\{\psi_{j},\psi_{l}\} = \delta_{jl}$, and $\psi_{2N+1} = -\psi_1$. The Hamiltonian is equivalent to the transverse field Ising model under the Jordan-Wigner transformation. Yet, a local channel in the spin representation may not be a local channel in the fermion representation. 

Here we use the covariance matrix technique to simulate the amplitude damping channel, which is a Guassian channel. The covariance matrix, defined by
\begin{equation}
    M_{jl} = -i \tr((\psi_j \psi_l - \delta_{jl})\rho),
\end{equation}
fully characterizes a Guassian state $\rho$. One may introduce complex fermion annihilation and creation operators $c_j = (\psi_{2j}-i\psi_{2j+1})/\sqrt{2}, c^{\dagger}_j = (\psi_{2j}+i\psi_{2j+1})/\sqrt{2}$ and the number operator $c^{\dagger}_j c_j$. We will shortly denote the operators for the system fermions as $c_{j,S}$ and $c^{\dagger}_{j,S}$, where $S$ means ``system". The amplitude damping channel $\mathcal{A}_p$ for the $j$-th complex fermion can be realized by 
\begin{equation}
    \mathcal{A}^{[j]}_p (\rho_S) = \tr_A[U^{[j]}_{p,SA} (\rho_S \otimes |0_j\rangle_A\langle 0_j|) U^{[j]\dagger}_{p,SA}],
\end{equation}
where we have introduced auxiliary complex fermions with creation and annihilation operators $c^{\dagger}_{j,A}, ~c_{j,A}$ labelled by the site $j$, the state $|0_j\rangle_A$ is the ground state of the number operator $c^{\dagger}_{j,A} c_{j,A}$ of the ancilla, and 
\begin{equation}
    U^{[j]}_{p,SA} = e^{\theta (c^{\dagger}_{j,S} c_{j,A} - c^{\dagger}_{j,A} c_{j,S}) }, ~~ p = \sin^2 \theta.
\end{equation}
Such representation makes it clear that the amplitude damping channel is a Gaussian channel. The channel on the total system is a product of single-fermion channels, $\mathcal{N}_p = \otimes_j \mathcal{A}^{[j]}_{p}$. The covariance matrix $M_p$ of the decohered state $\mathcal{N}_p(\rho)$ is
\begin{equation}
    M_p = (1-p)M + pM^{(0)},
\end{equation}
where $M$ is the covariance matrix of $\rho$ and $M^{(0)}$ is the covariance matrix of $|0\rangle^{\otimes L}$ with nonzero entries $M^{(0)}_{2i-1,2i} = -M^{(0)}_{2i,2i-1} = 1$. Using the covariance matrix, one may compute all the Renyi mutual information $I^{(n)}(A,B)$ and Renyi negativity $\mathcal{E}^{(n)}(A,B)$ of any two subsystems. One can directly take the replica limit to obtain the Von-Neumann mutual information $I(A,B)$ and the logarithmic negativity $\mathcal{E}(A,B)$ of the decohered state. We note that here the negativity is defined through partial time reversal rather than partial transpose, following the computation of Refs.~\cite{Shapourian_2017_negativity,Liu_2022_vertex}. The two definitions of negativity have different quantum information properties \cite{Shapourian2019PRA} as well as different path integral representations \cite{Shapourian2019scipost,Murciano2022}. The mapping to boundary CFT (Eq. (5)) of the main text only strictly applies to the definition through partial transpose. The mapping does not strictly apply to the free fermion case considered in this appendix as we are using the other derinition of negativity. Yet, we believe that the negativity considered here still maps to a two-point correlation function of boundary condition changing operators, although the boundary conditions are different from $|\mathcal{B}_{\mathcal{N}}\rangle\rangle$ and $|\tilde{\mathcal{B}}_{\mathcal{N}}\rangle\rangle$ in the main text.

As shown above, we find that the amplitude damping channel forms a continuous set of conformal fixed points with $\log g^{(n)}=0$ for any $n$. The scaling dimension of the boundary condition changing operator $\phi^{(n)}_{\mathcal{I}\mathcal{N}}$ also continuously changes with the damping strength $p$, as is evident by the (Renyi) mutual information $I^{(n)}(A,\bar{A})$. The (Renyi) negativity $\mathcal{E}^{(n)}(A,\bar{A})$ shows a similar logarithmic form with continuously changing coefficients, indicating a continuous set of conformal boundary conditions.  

\begin{figure}[h]
    \centering
    \includegraphics[width = 0.55\textwidth]{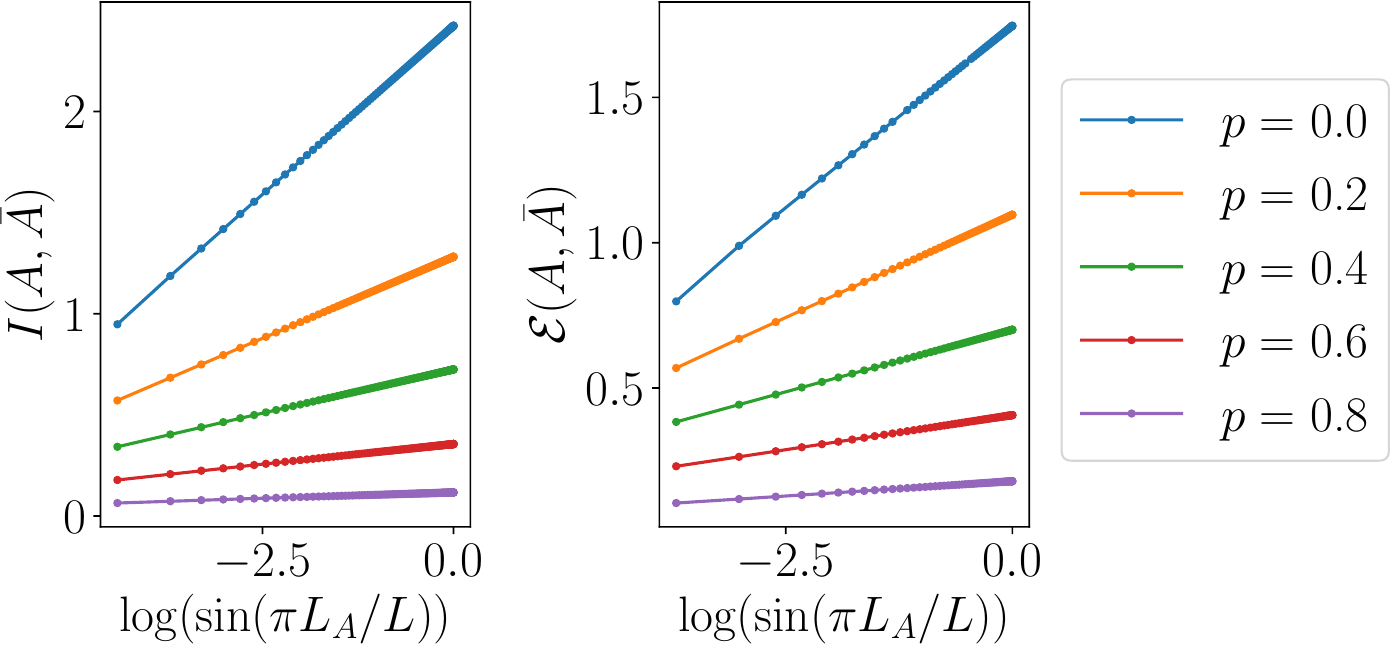}
    \caption{Mutual information (left) and negativity (right) for free fermion critical state subject to amplitude damping channel of strength $p$.}
    \label{fig:ff}
\end{figure}

In the main text we have found that the $Z$-dephasing also constitutes a continuous family of conformal fixed points with continuously varying coefficient $\Delta^{(n)}_{NT}$. One may wonder whether it may be related to the amplitude damping channel of the free fermion. Although the two sets of channels share some similarities, We note two important differences. Firstly, although the $Z$-depahsing under Jordan-Wigner transformation is still a local channel of the fermions, it is not a Gaussuan channel as opposed to the ampltude damping channel. Thus it is not straightforward to apply covariance matrix techniques to simulate the dephasing channel. Secondly and more importantly, the coefficient in front of the logarithm in the mutual information $I(A,\bar{A})$ are different. For the $Z$-dephasing, the scaling dimension $\Delta^{(n)}_{\mathcal{I}\mathcal{N}}$ does not change with $p$, whereas the same scaling dimension changes continuously with $p$ for the amplitude damping channel. Intuitively, thus makes sense as the amplitude damping channel tries to damp the system towards a product state with no classical correlations, whereas the dephasing channel still keeps the classical correlations along the $Z$ direction. It would be interesting to study whether the two sets of conformal boundary conditions can be related. This would require the identification with conformal boundary conditions for the multi-copied free fermion CFT, which we leave to future work.

\section*{Analytical treatment of Renyi mutual information}
Here we present analytical result of Renyi mutual information $I^{(n)}(A:B)$, where $A$ and $B$ are of $O(1)$ size. We show that $I^{(n)}(A:B)$ decays as a power law of the distance and the power is determined by the quantum channel. For the case of dephasing channel and $|A|,|B|=1$, we further take the replica limit $n\rightarrow 1$ and consider the Von-Neumann mutual information. Our results provide a concrete example where any Renyi mutual information with $n>1$ is not monotonic under local quantum channels, in sharp constrast with the Von-Neumann mutual information. 

\subsection{Renyi mutual information of two sites}
Let $|A| = |B| = 1$, then 
\begin{equation}
    I^{(n)}(A:B) = \frac{1}{n-1} \log \frac{\langle \mathcal{N}^{*}(\tau_{n,A})  \mathcal{N}^{*}(\tau_{n,B})\rangle}{\langle  \mathcal{N}^{*}(\tau_{n,A})\rangle \langle  \mathcal{N}^{*}(\tau_{n,B})\rangle} ,
\end{equation}
where the expectation value is taken on the ground state of $n$-copied CFT. Let the distance of $A$ and $B$ be $l$ and let 
\begin{equation}
    d = \frac{l}{\pi} \sin \frac{\pi l}{L}
\end{equation}
be the chord distance. We can expand the lattice operator $B^{(n)}_{\mathcal{N}} = \mathcal{N}^{*}(\tau_n)$ into a sum of CFT operators in $n$-copied CFT,
\begin{equation}
    B^{(n)}_{\mathcal{N}} = a_0I + a_1 O^{(n)} + \cdots, 
\end{equation}
where the $\cdots$ represents operators with higher scaling dimensions. The correlation function $\langle O^{(n)}(x) O^{(n)}(y) \rangle \sim d^{-2\Delta^{(n)}_{O}}$, thus
\begin{equation}
    I^{(n)}(A:B) \sim d^{-2\Delta^{(n)}_{O}}.
\end{equation}
\subsection{Renyi index $n=2$}
For complete dephasing, we compute the lattice operator and find their CFT counterpart using the dictionary in Ref.~\cite{Zou_2020_conformal}. Below we list results for various choices of $\vec{v}$ in the dephasing channel. For the fixed points $\mathcal{D}_z$, $\mathcal{D}_x$, $\mathcal{D}_{zx}$, we find that the leading CFT operator to $O^{(2)}$ is $\varepsilon \otimes I + I \otimes \varepsilon$, $\sigma \otimes \sigma$ and $\sigma \otimes I + I \otimes \sigma$, with scaling dimension $1,1/4,1/8$ respectively. This agrees with the numerical result shown in the main text, $I^{(2)}(A,B) \sim \eta^{\Delta^{(n)}_{O}}$. However, we note an important difference: in the main text we consider $|A|$ and $|B|$ to be much larger than the lattice spacing rather than $|A|=|B|=1$. It is not a priori the case that the scaling operators $O^{(n)}$ are the same for the two cases. Yet for the dephasing fixed points we observe that they coincide.

We also note that the exponent $1/8$ for $\mathcal{D}_{zx}$ is smaller than that of the identity channel, meaning that the channel increases the Renyi mutual information. As we show shortly, this happens for all Renyi indices $n>1$, but ceases to hold at $n=1$.
\begin{table}[h]
    \centering
    \begin{tabular}{|c|c|c|c|}
    \hline
        Dephasing direction $\vec{v}$ & $O^{(2)}$ (leading) & CFT operator & $\Delta^{(2)}_{O}$ \\ \hline
         $\vec{x}$ &  $X\otimes X$ & $\sigma \otimes \sigma$ & $1/4$ \\ \hline
         $\vec{y}$ &  $Y\otimes Y$ & $\partial_\tau\sigma \otimes \partial_{\tau}\sigma$ & $9/4$ \\ \hline
         $\vec{z}$ &  $Z\otimes Z$ & $\varepsilon \otimes I + I \otimes \varepsilon$ & $1$ \\ \hline
         Generic $(v_x,v_y,0)$ &  $X\otimes X$ & $\sigma \otimes  \sigma$ & $1/4$ \\ \hline
         Generic $(v_x,0,v_z)$ &  $X\otimes Z + Z\otimes X$ & $\sigma \otimes I + I \otimes \sigma$ & $1/8$  \\ \hline
         Generic $(0,v_y,v_z)$ &  $Z\otimes Z$ & $\varepsilon \otimes I + I \otimes \varepsilon$ & $1$ \\ \hline
         Generic $(v_x,v_y,v_z)$ &  $X\otimes Z + Z \otimes X$ & $\sigma \otimes I + I \otimes \sigma$ & $1/8$ \\ \hline
    \end{tabular}
    \caption{Scaling dimension of $O^{(2)}$ under complete dephasing ($p=1$) in different directions. (``Generic" means all components are nonzero.)}
    \label{tab:Ising_dephasing1}
\end{table}

\subsection{Higher Renyi index}
For general replica number $n\geq 2$, we first consider the $\vec{z}$ direction dephasing,
\begin{equation}
    B^{(n)}_{\mathcal{N}}:=\mathcal{D}^{* \otimes n}_{p=1,\vec{z}}(\tau_n) = \sum_{i=0}^{1} | i^{\otimes n}\rangle \langle i^{\otimes n}|
\end{equation}
In Pauli basis, this is
\begin{equation}
    \mathcal{D}^{* \otimes n}_{p=1,\vec{z}}(\tau_n) = \frac{1}{2^{n-1}}\sum_{s\in S_{\vec{z}}} s,
\end{equation}
where $S_{\vec{z}}$ is the stabilizer group consisting of even number of product of Pauli $Z$'s. Note that this expression also holds if we change $z$ to any $\vec{v}$ and correspondingly the stabilizer group $S_{\vec{z}}$ to $S_{\vec{v}}$ (which is isomorphic to $S_{\vec{z}}$ by changing $Z$ to $\vec{v}\cdot \vec{\sigma}$). Now we consider the correlation function of this operator on $A$ and $B$. Taking into account that the correlation function of a single Pauli string factorizes into product of correlation functions of one replica, we obtain
\begin{eqnarray}
    \langle \psi^{\otimes n}|B^{(n)}_{\mathcal{N}}(x)B^{(n)}_{\mathcal{N}}(y)|\psi^{\otimes n} \rangle &=& ((1+x)^2+C)^n + ((1-x)^2+C)^n + 2 (1-x^2-C)^n \\
    \langle \psi^{\otimes n}|B^{(n)}_{\mathcal{N}}(x)|\psi^{\otimes n} \rangle \langle \psi^{\otimes n}|B^{(n)}_{\mathcal{N}}(y)|\psi^{\otimes n} \rangle &=& (1+x)^{2n} + (1-x)^{2n} +2(1-x^2)^n
\end{eqnarray}
where $x= \langle\sigma_{\vec{v}}\rangle, C= \langle \sigma_{\vec{v}}(x) \sigma_{\vec{v}}(y)\rangle_c$ is the connected correlation function of the ground state, and $\sigma_{\vec{v}} \equiv \vec{v} \cdot \vec{\sigma}$. The Renyi mutual information is given by
\begin{equation}
    I^{(n)}(A,B) = \frac{1}{n-1} \log \frac{\langle \psi^{\otimes n}|B^{(n)}_{\mathcal{N}}(x)B^{(n)}_{\mathcal{N}}(y)|\psi^{\otimes n} \rangle}{\langle \psi^{\otimes n}|B^{(n)}_{\mathcal{N}}(x)|\psi^{\otimes n} \rangle \langle \psi^{\otimes n}|B^{(n)}_{\mathcal{N}}(y)|\psi^{\otimes n}\rangle}.
\end{equation}
In the case of $n=2$, this reduces to
\begin{equation}
    I^{(2)}(A,B)= \log \left(1+\frac{x^2 C +C^2}{(1+x^2)^2}\right)
\end{equation}
In the limit where $C\ll 1$, we obtain
\begin{equation}
    I^{(2)}(A,B) = \frac{x^2 C +C^2}{(1+x^2)^2}~~ (C\ll 1)
\end{equation}
This is in agreement with our previous subsections. If $x=0$, that is, if the dephasing direction has $v_x = 0$, then $I^{(2)}$ is proportional to the square of correlation function. However, if $x\neq 0$, then $I^{(2)}$ gets promoted to linear in the correlation function. This means that dephasing can actually increase the second Renyi mutual information (e.g., for $\mathcal{D}_{zx}$). More generally, if $n\geq 2$, we have
\begin{equation}
    I^{(n)}(A,B)= \frac{1}{n-1}\log \left(1+\frac{n((1+x)^{n-1}-(1-x)^{n-1})^2C + O(C^2) }{((1+x)^n+ (1-x)^n)^2}\right),
\end{equation}
which reduces to 
\begin{equation}
    I^{(n)}(A,B)= \frac{n}{n-1}\frac{((1+x)^{n-1}-(1-x)^{n-1})^2}{((1+x)^n+ (1-x)^n)^2} C+ O(C^2), ~~~(n\geq 2, C\ll 1)
\end{equation}

Thus, for all Renyi mutual information $I^{(n)}(A,B)$ with $n>1$, it can be promoted to linear in correlation function by acting with a dephasing channel $\mathcal{D}_{p=1,\vec{v}}$ if $C:=\langle \sigma_{\vec{v}}\rangle \neq 0$.

\subsection{Replica limit $n\rightarrow 1$}
One important point is that the Von Neumann mutual information cannot increase under noise channels. To see this, we use the data processing inequality in quantum information theory
\begin{equation}
    S(\mathcal{N}(\rho)|\mathcal{N(\sigma)})\leq S(\rho|\sigma)
\end{equation}
Using the fact that $I(A,B) = S(\rho_{AB}|\rho_{A}\otimes \rho_B)$ and the fact that $\mathcal{N} = \mathcal{N}_A\otimes \mathcal{N}_B$, we obtain
\begin{equation}
    I(A,B)_{\mathcal{N}(\rho)}\leq I(A,B)_{\rho}.
\end{equation}
The data processing inequality does not apply to the Renyi entropies, so the Renyi mutual information is not monotonic under product channels.

The information-theoretic argument dictates that the Von Neumann mutual information after applying the noise channel should be at least quadratic in the original correlation functions. In order to see this, we take the replica limit $n\rightarrow 1$. The result is
\begin{equation}
    I(A,B) = \left(H\left(\frac{(1+x)^2+C}{4}\right) +H\left(\frac{(1-x)^2+C}{4}\right) + 2H\left(\frac{1-x^2-C}{4}\right)\right) - (C \leftarrow 0)
\end{equation}
where $H(x) = x \log x$. Note that the approximation with $C\ll 1$ must come \textit{after} taking the replica limit. Using the Taylor expansion around $C=0$, and the derivatives $H'(x) = 1+\log x,~H''(x) = 1/x$, we obtain
\begin{equation}
    I(A,B) = \frac{C^2}{2(1-x^2)^2} ~~(C\ll 1).
\end{equation}
Note that the linear term in $C$ exactly cancels. Thus we expect that
\begin{equation}
    I(A,B) = O(d^{-4\Delta_{\vec{v}\cdot \vec{\sigma}}})
\end{equation}
under complete dephasing. See the table below for a list of scaling dimensions for different dephasing channels. We no longer observe an increase in the Von-Neumann mutual information, as required by the data processing inequality.
\begin{table}[h]
    \centering
    \begin{tabular}{|c|c|c|c|}
    \hline
        Dephasing direction $\vec{v}$ & $\vec{v}\cdot \vec{\sigma}$ (leading) & CFT operator & $4\Delta_{\vec{v}\cdot \vec{\sigma}}$ \\ \hline
         $\vec{x}$ &  $X$ & $\sigma$ & $1/2$ \\ \hline
         $\vec{y}$ &  $Y$ & $\partial_\tau\sigma$ & $9/2$ \\ \hline
         $\vec{z}$ &  $Z$ & $\varepsilon$ & $4$ \\ \hline
         Generic $(v_x,v_y,0)$ &  $X$ & $\sigma$ & $1/2$ \\ \hline
         Generic $(v_x,0,v_z)$ &  $X$ & $\sigma$ & $1/2$  \\ \hline
         Generic $(0,v_y,v_z)$ &  $Z$ & $\varepsilon$ & $4$ \\ \hline
         Generic $(v_x,v_y,v_z)$ &  $X$ & $\sigma$ & $1/2$ \\ \hline
    \end{tabular}
    \caption{Mutual information scaling under complete dephasing ($p=1$) in different directions.}
    \label{tab:Ising_dephasing3}
\end{table}
\section*{Applicability of conformal boundary conditions}
In a global quantum quench problem to a critical Hamiltonian, not all short-range entangled initial state flows to a conformal boundary state. Yet, it has been argued by Cardy \cite{Cardy_2017_bulk} that the ground state of the CFT Hamiltonian perturbed by a relevant deformation flows to a Cardy state, a conformal boundary state subject to physical constraints. 

In this appendix, we show that the state $|B_{\mathcal{N}}\rangle\rangle$ for a dephasing channel considered in the main text can be associated with such relevant deformations to the $2n$-copied Ising model. Thus, if the above Cardy's conjecture holds, we can argue that $|B_{\mathcal{N}}\rangle\rangle$ in the main text all flows to conformal boundary states. We also provide an example of $\mathcal{N}$ such that the argument fails and such that $|B_{\mathcal{N}}\rangle\rangle$ does not flow to a conformal boundary state. We will stay within replica index $n=2$ in this appendix.

To start with, we consider the case where $\mathcal{N} = \mathcal{D}_{p=1,x}$. In this case the operator
\begin{equation}
    B^{(2)}_{\mathcal{N}}:= \mathcal{N}^{*\otimes 2}(\tau_2) = \frac{1}{2}(I + X\otimes X)
\end{equation}
Under the folding, the operator becomes a state in 4 replicas, which is
\begin{equation}
    |B^{(2)}_{\mathcal{N}}\rangle\rangle = \frac{1}{2}(|++++\rangle + |----\rangle),
\end{equation}
where $|+\rangle$ or $|-\rangle$ are the eigenstates of $X$ in one replica. One parent Hamiltonian of $|B^{(2)}_{\mathcal{N}}\rangle\rangle$ is 
\begin{equation}
    H^{(2)}_{int,\mathcal{N}} = X \otimes X \otimes I \otimes I +  I \otimes X \otimes X \otimes I + I \otimes I \otimes X \otimes X.
\end{equation}
The state $|B^{(2)}_{\mathcal{N}}\rangle\rangle$ is its unique ground state upon fixing the total parity. Note that this Hamiltonian is defined on \textit{one} lattice site across the four replicas. Now we consider the replicated CFT Hamiltonian
\begin{equation}
    H^{(2)}_{C} = H_{C}\otimes I \otimes I \otimes I + I\otimes H_{C} \otimes I \otimes I + I\otimes I \otimes H_{C} \otimes I + I\otimes I \otimes I \otimes H_{C},
\end{equation}
where $H_{C}$ is the Hamiltonian for one copy of critical Ising model. Now we consider the perturbation of $H^{(2)}_C$ by the parent Hamiltonian,
\begin{equation}
    H^{(2)} = H^{(2)}_{C} +\lambda \sum_{j=1}^L H^{(2)}_{int,\mathcal{N},j} 
\end{equation}
The interaction term $H^{(2)}_{int,\mathcal{N}}$ is a \textit{relevant} deformation since its scaling dimension is $2\Delta_{\sigma} = 1/4<2$. At long wavelengths, the coupling $\lambda$ flows to infinity, and the ground state of $H^{(2)}$ is qualitatively the same as the ground state of $H^{(2)}_{int,\mathcal{N}}$, which is $|B^{(2)}_{\mathcal{N}}\rangle\rangle$. Thus, if the aforementioned Cardy's conjecture holds, then $|B^{(2)}_{\mathcal{N}}\rangle\rangle$ can be described by a conformal boundary state. 

In order to generalize the argument to other dephasing channels, we just have to replace $X$ in the parent Hamiltonian $H^{(2)}_{int,\mathcal{N}}$ by $\sigma_{\vec{v}} = \vec{v}\cdot \vec{\sigma}$. If $\vec{v}$ is in the $XZ$ plane, then all terms in the parent Hamiltonian are relevant (they are either $\sigma\otimes \sigma$, $\sigma\otimes I$ or $\sigma \otimes \varepsilon$), and the argument above goes through. We note in passing that for all Renyi index $n\geq 2$ the argument also holds. However, if $\vec{v}$ is in the $YZ$ plane, then not all terms are relevant. For example, the term $Y\otimes Y \otimes I \otimes I$ has a scaling dimension $2\Delta_{\partial\sigma} = 9/4>2$. Then it is no longer guaranteed that the boundary state $|B^{(2)}_{\mathcal{N}}\rangle\rangle$ flows to a conformal boundary condition.

The applicability of conformal boundary condition is crucial to all the main results. The main results could fail, if the boundary state $B^{(n)}_{\mathcal{N}}$ does not flow to a conformal boundary state. Indeed, if we choose $\vec{v} = (0, \sin \theta, \cos \theta)$ in the $YZ$ plane, then we observe that the $g$-function is monotonically increasing rather than decreasing for a range of $\theta$, e.g., $0<\theta<\pi/4$. Yet, the $g$-function still appears to converge at large $L$, indicating scale invariance. These boundary conditions may be examples of scale invariant but not conformal boundary conditions in the replicated CFT. We leave it as future work to explore these boundary conditions.

\section*{Numerical techniques}
In this appendix, we describe matrix product state techniques to evaluate $S^{(n)}_A(\mathcal{N}(\ket{\psi}\bra{\psi}))$. For the sake of presentation, all the tensor diagrams were drawn for the special case that $n=2$, $L=4$ and $A=\{1,2\}$, however the generalization to generic cases is straightforward.

By performing density matrix renormalization group (DMRG) simulation, one is able to obtain the matrix product state (MPS) representation of the critical ground state $\ket{\psi}$ of interest:
\begin{equation}
    \ket{\psi} = \eqfig{1.cm}{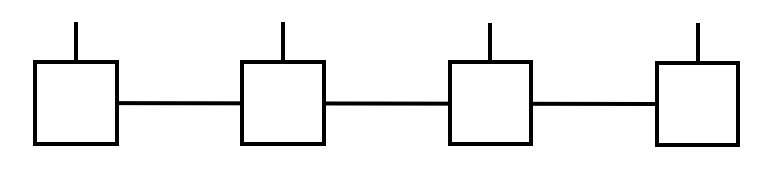}
\end{equation}
Accordingly, the partition function $Z^{(n)}_A$ can be diagrammatically represented as:
\begin{equation}
    Z^{(n=2)}_{A=\{1,2\}} = \eqfig{5.cm}{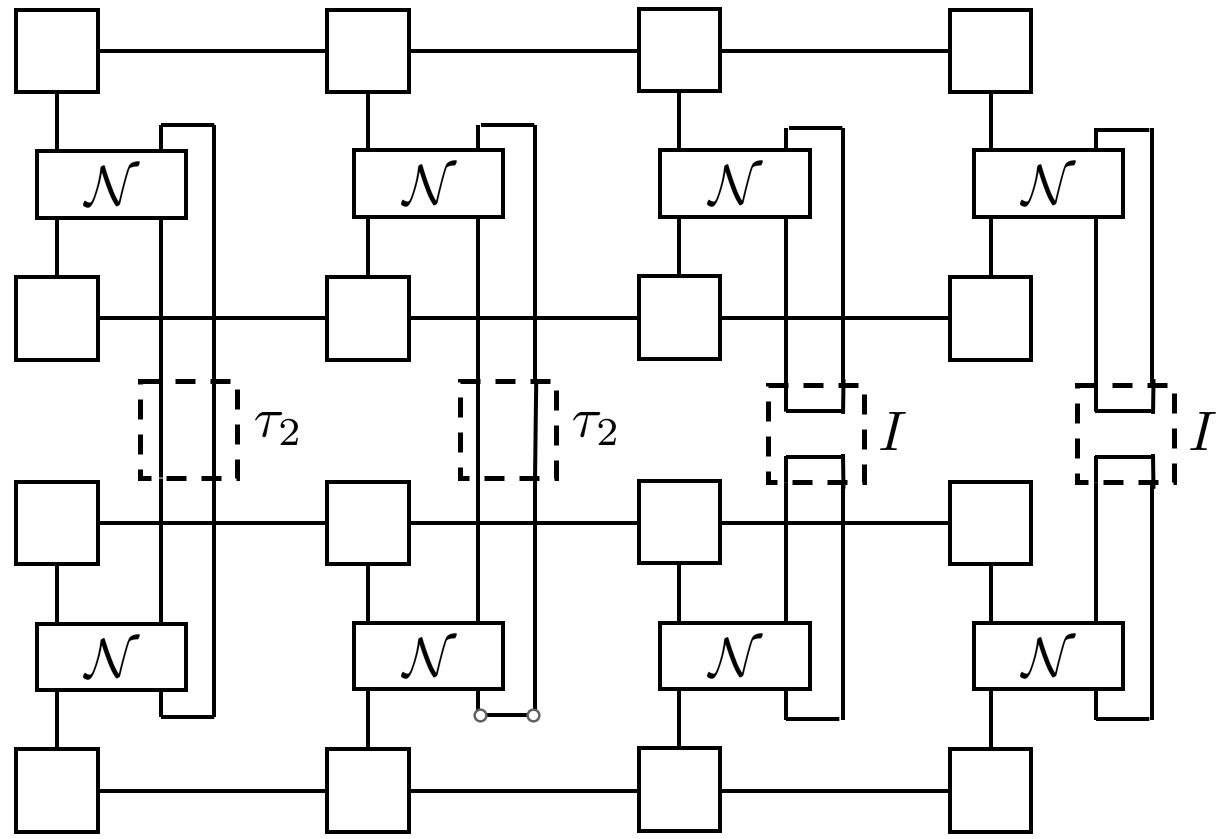}
\end{equation}
By contracting from left to right, the diagram can be contracted with a time complexity $O(L\chi^{2n+1})$, where $\chi$ is the maximum bond dimension of the MPS. Using the above techniques, we can evaluate $Z^{(2)}$ for $\chi$ up to $100$ and $L$ up to $256$.

By replacing $I$ with $\tau^{-1}_n$, the same technique can also be used for evaluating Renyi entanglement negativity.

\section{Experimental protocol}
Here we propose an experimental protocol to simulate the proposed phase transitions among noisy channels. Recall that we classify noisy channels into relevant, irrelevant and marginal depending on different scaling of the Renyi negativity. Our proposal consists of the following five steps.
\begin{itemize}
    \item 1) Preparing a critical ground state of $L$ spins on a programmable quantum computer such as an ion trap.
    \item 2) Apply decoherence channel $\mathcal{N}$ on each qubit.
    \item 3) Doing projective measurements on each qubit on a random basis $\{U_i|0\rangle, U_i|1\rangle\}$, where $U_i$ is a site-dependent Haar random $2\times 2$ unitary matrix. This output a bit string of measurement outcomes $b \in \{0,1\}^{\otimes L}$.
    \item 4) Repeat steps 1)-3) $N$ times, which gives a sample of $N$ bitstrings $\{b\}$ labeled by the known random unitary $\{U_i\}$.
    \item 5) Using classical computation to obtain the Renyi negativity $N^{(3)}_A$ from the set $\{b, U_i\}$ for all the choices of subsystem $A$. 
\end{itemize}
Several steps have been achieved in previous works. For step 1), the ground state can be prepared either using the hybrid quantum-classical energy optimization (e.g., QAOA) method as in Ref.~\cite{zhuGeneration2020a} or using a MERA circuit as in Ref.~\cite{anand2022holographic}. Both works prepare the critical ground state on an ion trap quantum computer. 
The critical state preparation should be as fast as possible, such that the decoherence introduced is small. Fortunately, neither the QAOA nor MERA has a long circuit depth. In the former case the depth is proportional to $L$. In the latter case, only $O(\log L)$ depth and a total number of $O(L)$ gates are needed. We will comment on the effect of errors in the state preparation process shortly.

The step 2) introduces one-shot decoherence $\mathcal{N}^{\otimes L}$, which is the central object of our study. The channel can be implemented by one of the three following ways. (1) We manually couple the system qubit to an ancillary qubit and then discard the ancilla. The coupling is chosen such that they constitute a unitary manifestation of the channels of interest; (2) We apply the Kraus operator with some probability but forget about the choice, e.g., for a $X$-dephasing channel $\mathcal{D}_{p,X}$, we apply the bit flip with probability $p/2$; (3) For realistic systems where one type of decoherence dominates, we can apply a local unitary change of basis and wait for the environment to decohere the system.

\begin{figure}
    \centering
    \includegraphics[width = 0.9 \linewidth]{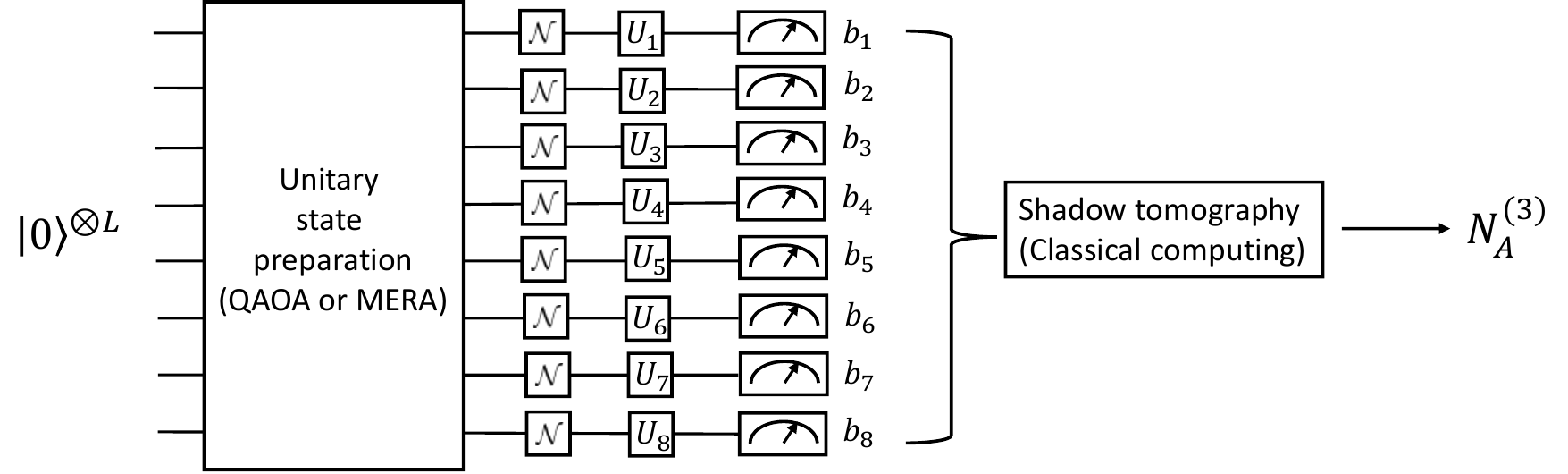}
    \caption{Experimental protocol for computing the Renyi negativity of a decohered critical state}
    \label{fig:exp}
\end{figure}

The steps 3)-5) are known as the shadow tomography method. The method utilizes random measurements to reconstruct properties of the quantum state. It is known how to compute the Renyi entanglement entropy and negativity from the classical shadows, see Refs.~\cite{Huang_2020_predicting,Elben_2020_mixed} for details. The benefit of using shadow tomography is that one can predict many different quantities (in this case, Renyi negativity of different subsystems $A$) with the same data set obtained in step 4). The error of estimation decreases as $N^{-1/2}$ with the number of samples $N$. We will assume that we have enough samples to perform an accurate tomography of the entanglement quantity of interest. We remark that the number of samples need to grow exponentially with $L$ for a given accuracy of estimation of Renyi entropy or Renyi negativity. Yet, on an intermediate-scale quantum device, it is still feasible to obtain a reasonably accurate estimation \cite{Huang_2020_predicting,Elben_2020_mixed}. The whole procedure is summarized in Fig.~\ref{fig:exp}. 

The prediction of our theory is that if the decoherence channel $\mathcal{N}$ corresponds to an irrelevant or marginal channel, the Renyi negativity would still be logarithmic in the chord length of subsystem $A$, $\sin(\pi L_A/L)$. On the other hand, if the decoherence corresponds to a relevant channel, the Renyi negativity would become area law.

We stress that our theory of RG flow in quantum channels only strictly applies to the decoherence introduced in step 2). The decoherence introduced in the state preparation would result in long-ranged correlated noise which is beyond the scope of the current study. However, we argue that a slight noise introduced in the state preparation process would not change the prediction of negativity scaling if the prepared state still have significant overlap with the low-energy subspace. From the mapping to the boundary CFT in the main text, the RG flow between quantum channels is a property of the whole low-energy subspace rather than just the ground state. The full consequence of this observation will be explored in a follow-up work. One immediate consequence is that the channel $\mathcal{N}$ is described by the same fixed point in the whole low-energy subspace. Thus, starting from any low-energy states of a critical Ising model which could be possibly mixed, the $X$-dephasing always drives the system to an area law state. On the other hand, the $Z$-dephasing, being a marginal channel on the low-energy subspace, retains long-range quantum entanglement and the resulting state would have a logarithmic scaling of (Renyi) entanglement negativity. Furthermore, we make a qualitative estimate of the rate of noise $p_e$ per circuit depth which this argument applies in the case of QAOA. The low-energy subspace is characterized by zero energy density. Thus, (1) if the noise rate is less than $O(1/L^2)$ and the circuit depth is $O(L)$, then the noise is not sufficient to move the state out of the low-energy subspace and our argument applies. In this case we can observe the theoretical prediction perfectly. (2) If the noise rate is larger but still less than $O(1/L)$, then generically the prepared state can still support nontrivial quantumness, as the time scale (depth) where thermalization happens is $O(1/p_e) \geq O(L)$ \cite{Li2022Entanglement}. In this case we can observe some signatures that are qualitatively similar to our theoretical prediction. (3) If the noise rate is greater than $O(1/L)$, then the prepared state trivializes. In the following simulation, we see that on a current stage NISQ device, say, $L=16$ spins with $p_e\approx 10^{-2}$, the first case can be achieved. This means that our theoretical prediction can be perfectly reproduced in a current-stage quantum simulator.

\section{Simulation of the experimental protocol}

\subsubsection{Case 1: Device does not have a preferred noise}
To corroborate our argument above, we perform a classical simulation of our experimental protocol on $L=16$ spins. We prepare the critical ground state using the classically optimized QAOA circuit \cite{10.21468/SciPostPhys.6.3.029}, where we incorporate a rate $p_e$ of depolarization noise for all two-qubit gates. The noise rate of single-qubit gates on an actual quantum simulator is much smaller and we neglect it for simplicity. In the simulation we consider three different noise rates, $p_e = 0.01, 0.03, 0.05$. We then subject the prepared noisy state to dephasing channels $D_{p,\vec{v}}$, where we fix $p = 0.3$ and $\vec{v} =x,y,z$. Note that we should distinguish $p_e$ and $p$, where the former one should be avoided in the ideal case and the latter one is a tuning parameter of our theory. 

\begin{figure}
    \centering
    \includegraphics[width = 0.45\linewidth]{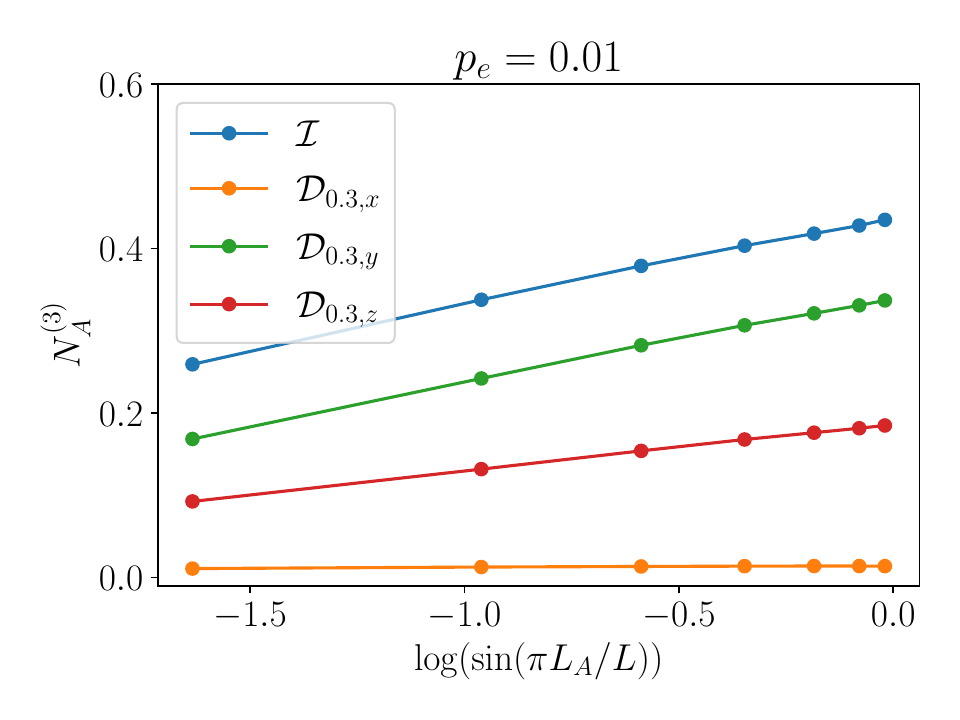}
    \includegraphics[width = 0.45\linewidth]{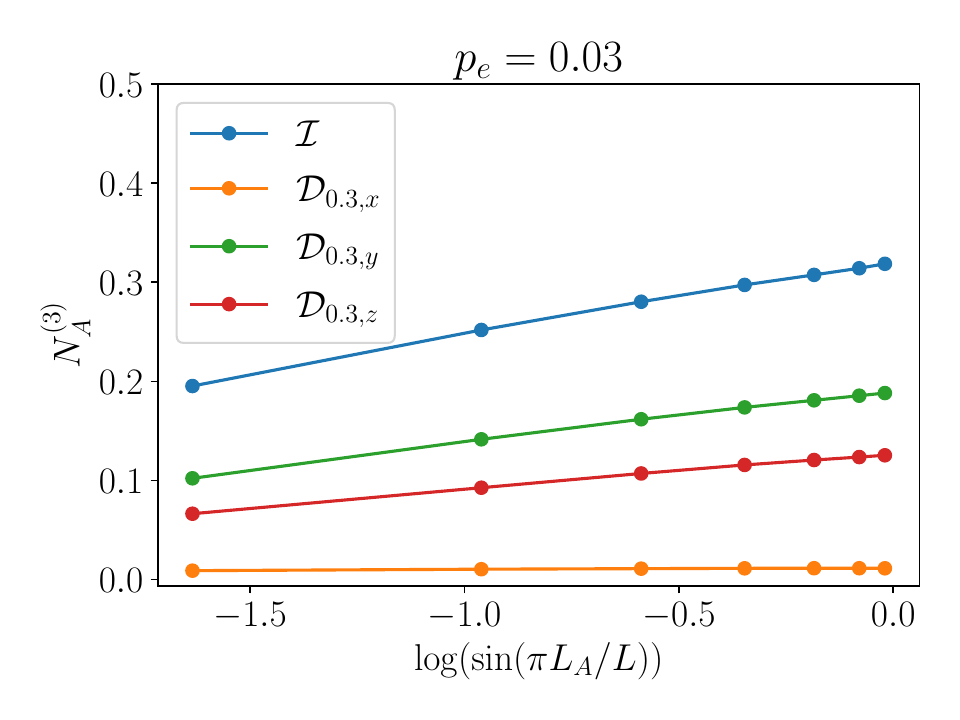}
    \includegraphics[width = 0.45\linewidth]{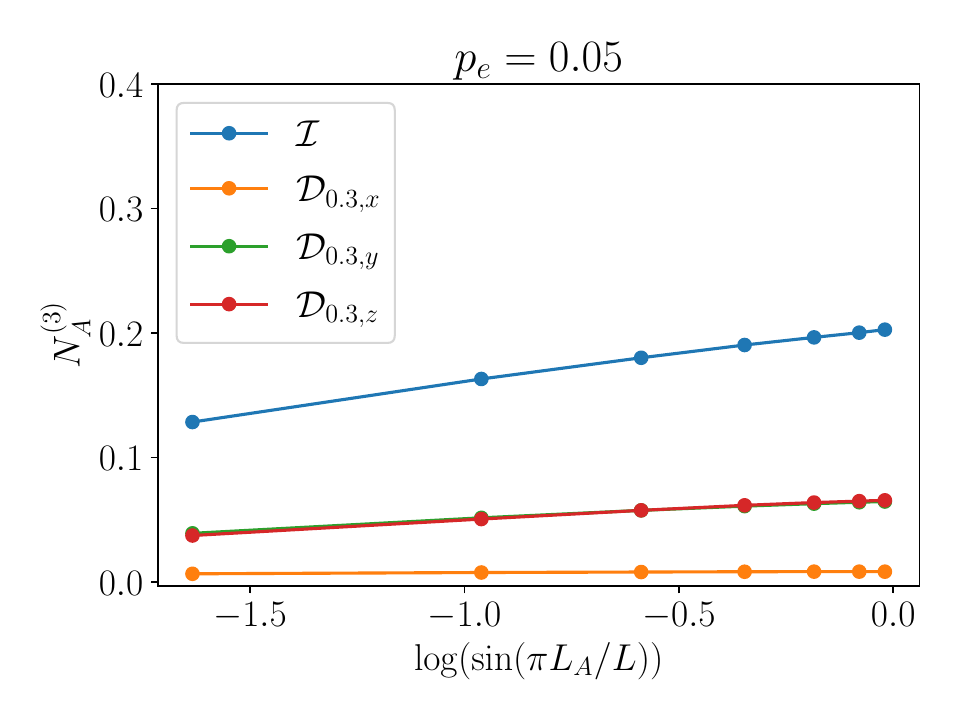}
    \caption{Simulation of the experimental protocol in an unbiased environment.}
    \label{fig:simulation_depo}
\end{figure}

Assuming that the shadow tomography of the Renyi negativity $N^{(3)}_A$ are accurate, we compute the slope $\Delta^{(3)}_{NT}$ in Eq.~\eqref{eq:NT3Y} by linearly fitting $N^{(3)}_A$ versus $\log(\sin (\pi L_A/L))$. The result is shown in Fig.~\ref{fig:simulation_depo}. We see that even at small system size and the noise rate $p_e = 0.01$, we still get a sharp distinction of $\Delta^{(3)}_{NT}$ for irrelevant/relevant/marginal dephasing channels. For the $Y$ dephasing, $\Delta^{(3)}_{NT}$ is almost the same as no dephasing. For $X$ dephasing, $\Delta^{(3)}_{NT}\approx 0$ and the negativity is area law. For $Z$ dephasing, the negativity still scales logarithmically and $\Delta^{(3)}_{NT}$ equals a $p$-dependent nonzero number. Given that $10^{-2}$ is the typical noise rate for two-qubit gates in the current-stage ion trap quantum simulator (e.g., for the Quantinuum ion trap, the noise rate $p_e\approx 7.9\times 10^{-3}$, as quoted from Ref.~\cite{anand2022holographic}), our protocol is achievable on current experimental devices. Naively, the noise in the state preparation would spread into correlated noise in the final state and the prepared state is far from the ground state. This is evident by the fact that the purity of the prepared state is only roughly $0.25$ in the numerical simulation. Yet, the distinction among relevant/irrelevant/marginal channels acting on this state is still manifest. 

For larger error rate $p_e$, we find that the prepared state with QAOA has an approximately logarithmic scaling of Renyi negativity with a smaller coefficient. We observe that even with $p_e=0.05$ it is not sufficient to trivialize the prepared state. Now we add a strong dephasing to the prepared state and compute the Renyi negativity. While the difference between the irrelevant channel $D_{p,y}$ and the marginal channel $D_{p,z}$ becomes blurred, we can still distinguish them from the relevant channel $D_{p,x}$, which results in a very small negativity. The above simulation raises the following interesting question. Given a circuit which prepares the critical ground state, can we still define the RG flow among dephasing channels in the presence of noise in the state preparation? Is there a critical noise rate which trivializes the state prepared? The above simulation indicates that for QAOA on a small system size one can still tolerate a moderate amount of noise in the state preparation. However, the QAOA has a depth that is linear in $L$, and thus we expect that for a larger system it only tolerates a smaller error rate. If we instead use MERA to prepare the critical state, the effect of noise could be very different. We leave a full analysis of this question to future work.

\subsubsection{Case 2: Device has a preferred noise}
Furthermore, we consider the case where the noise of the quantum simulator is highly biased. We consider the biased noise channels on a single site,
\begin{eqnarray}
    D_{p_x,p_y,p_z}(\rho) = \frac{p_x}{2} X\rho X + \frac{p_y}{2} Y\rho Y + \frac{p_z}{2} Z\rho Z + \left(1-\frac{p_x+p_y+p_z}{2}\right) \rho.
\end{eqnarray}
For concreteness, we consider a bias ratio 10, which means that $p_x = 10 p_y = 10 p_z$ for a $X$-biased noise. Similarly we define the $Y$-biased noise and $Z$-biased noise. Given a quantum simulator with a preferred direction of noise, we can apply a layer of single-site unitary to make the preferred noise in any direction. Furthermore, after preparing the critical state, we can just apply a layer of unitary change of basis and allow the system to evolve for a certain amount of time, and the environment will automatically implement the desired dephasing channel. 

\begin{figure}
    \centering
    \includegraphics[width = 0.45\linewidth]{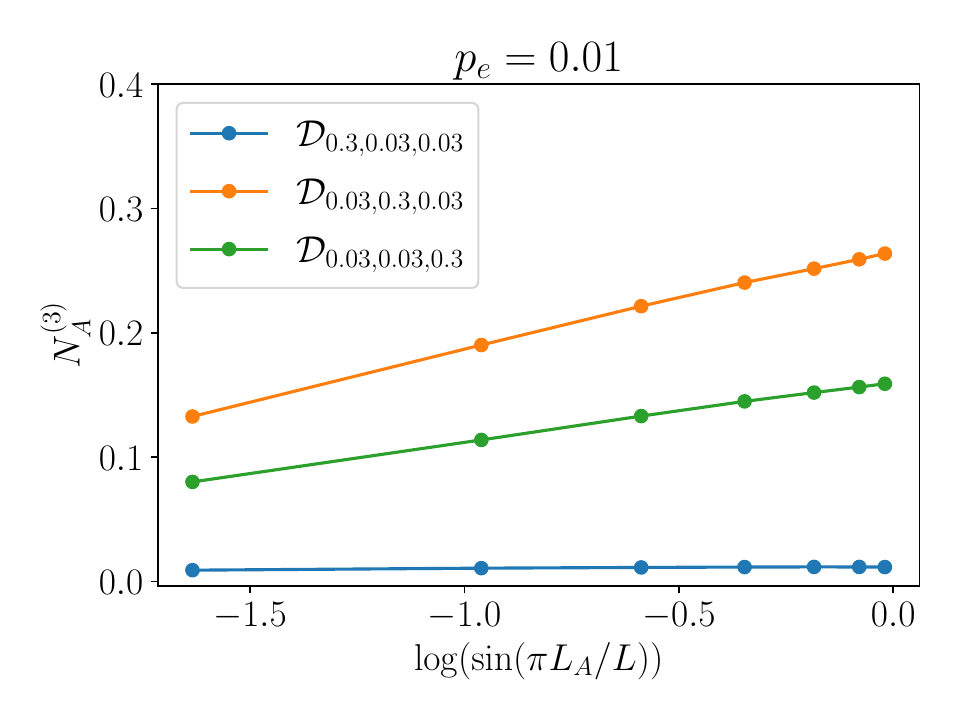}
    \includegraphics[width = 0.45\linewidth]{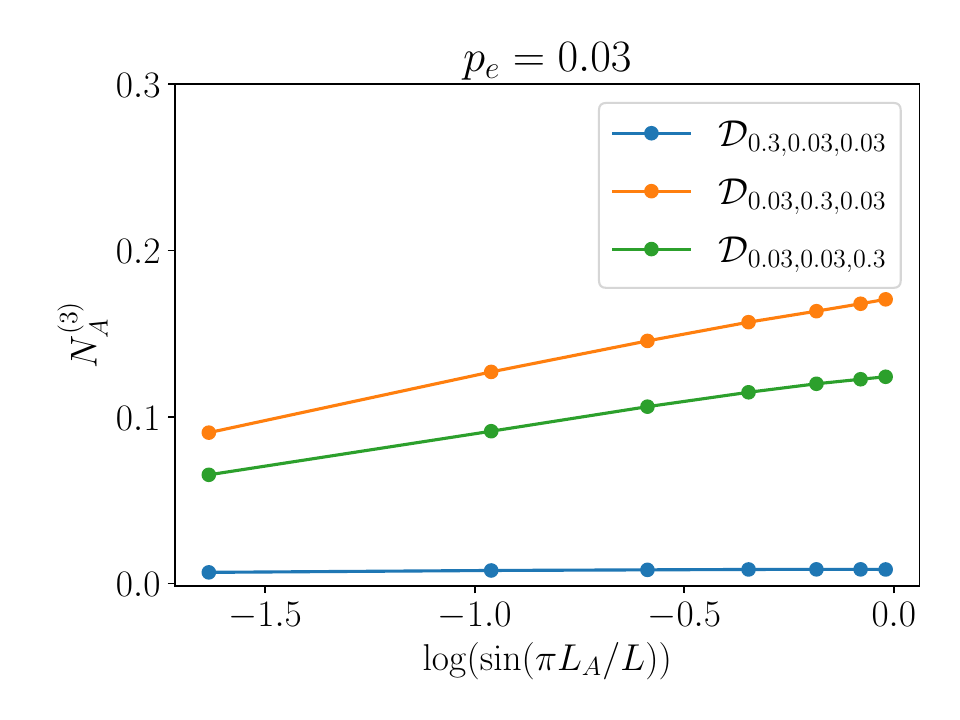}
    \includegraphics[width = 0.45\linewidth]{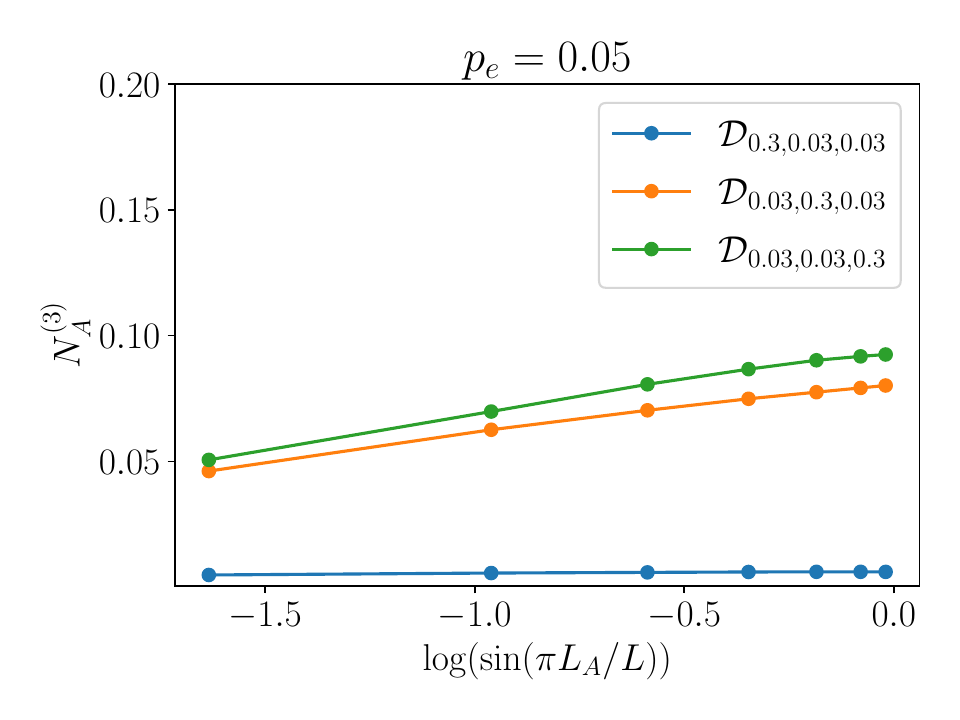}
    \caption{Simulation of the experimental protocol in a highly biased environment. The legend indicates the biased noise applied to the prepared state.}
    \label{fig:simulation_biased}
\end{figure}

In the simulation below, we consider the same QAOA circuit as in the last subsection. However, we replace the depolarization noise with the biased noise with rate in the preferred direction $p_e=0.01,0.03,0.05$. After the state preparation, we apply the same biased noise with a larger rate, which is fixed to $p=0.3$ in the preferred direction. Again, we find that there is a sharp distinction of the Renyi negativity among different biased noise, see Fig.~\ref{fig:simulation_biased}.

\end{document}